\PassOptionsToPackage{unicode}{hyperref}
\PassOptionsToPackage{hyphens}{url}
\PassOptionsToPackage{dvipsnames,svgnames,x11names}{xcolor}
\documentclass[
  11pt,
  a4paper,
]{article}

\usepackage{amsmath,amssymb}
\usepackage{iftex}
\ifPDFTeX
  \usepackage[T1]{fontenc}
  \usepackage[utf8]{inputenc}
  \usepackage{textcomp} 
\else 
  \usepackage{unicode-math}
  \defaultfontfeatures{Scale=MatchLowercase}
  \defaultfontfeatures[\rmfamily]{Ligatures=TeX,Scale=1}
\fi
\usepackage[]{libertinus}
\ifPDFTeX\else  
\fi
\IfFileExists{upquote.sty}{\usepackage{upquote}}{}
\IfFileExists{microtype.sty}{
  \usepackage[]{microtype}
  \UseMicrotypeSet[protrusion]{basicmath} 
}{}
\makeatletter
\@ifundefined{KOMAClassName}{
  \IfFileExists{parskip.sty}{%
    \usepackage{parskip}
  }{
    \setlength{\parindent}{0pt}
    \setlength{\parskip}{6pt plus 2pt minus 1pt}}
}{
  \KOMAoptions{parskip=half}}
\makeatother
\usepackage{xcolor}
\usepackage[a4paper,textheight=24cm,textwidth=15.5cm]{geometry}
\makeatletter
\ifx\paragraph\undefined\else
  \let\oldparagraph\paragraph
  \renewcommand{\paragraph}{
    \@ifstar
      \xxxParagraphStar
      \xxxParagraphNoStar
  }
  \newcommand{\xxxParagraphStar}[1]{\oldparagraph*{#1}\mbox{}}
  \newcommand{\xxxParagraphNoStar}[1]{\oldparagraph{#1}\mbox{}}
\fi
\ifx\subparagraph\undefined\else
  \let\oldsubparagraph\subparagraph
  \renewcommand{\subparagraph}{
    \@ifstar
      \xxxSubParagraphStar
      \xxxSubParagraphNoStar
  }
  \newcommand{\xxxSubParagraphStar}[1]{\oldsubparagraph*{#1}\mbox{}}
  \newcommand{\xxxSubParagraphNoStar}[1]{\oldsubparagraph{#1}\mbox{}}
\fi
\makeatother

\usepackage{color}
\usepackage{fancyvrb}

\DefineVerbatimEnvironment{Highlighting}{Verbatim}{commandchars=\\\{\}}
\newenvironment{Shaded}{}{}

\newcommand{\BuiltInTok}[1]{\textcolor[rgb]{0.84,0.23,0.29}{#1}}

\newcommand{\CommentTok}[1]{\textcolor[rgb]{0.42,0.45,0.49}{#1}}

\newcommand{\ControlFlowTok}[1]{\textcolor[rgb]{0.84,0.23,0.29}{#1}}

\newcommand{\DecValTok}[1]{\textcolor[rgb]{0.00,0.36,0.77}{#1}}

\newcommand{\FloatTok}[1]{\textcolor[rgb]{0.00,0.36,0.77}{#1}}

\newcommand{\KeywordTok}[1]{\textcolor[rgb]{0.84,0.23,0.29}{#1}}
\newcommand{\NormalTok}[1]{\textcolor[rgb]{0.14,0.16,0.18}{#1}}
\newcommand{\OperatorTok}[1]{\textcolor[rgb]{0.14,0.16,0.18}{#1}}

\newcommand{\StringTok}[1]{\textcolor[rgb]{0.01,0.18,0.38}{#1}}
\newcommand{\VariableTok}[1]{\textcolor[rgb]{0.89,0.38,0.04}{#1}}

\providecommand{\tightlist}{%
  \setlength{\itemsep}{0pt}\setlength{\parskip}{0pt}}\usepackage{longtable,booktabs,array}
\usepackage{calc} 
\usepackage{etoolbox}
\makeatletter
\patchcmd\longtable{\par}{\if@noskipsec\mbox{}\fi\par}{}{}
\makeatother
\IfFileExists{footnotehyper.sty}{\usepackage{footnotehyper}}{\usepackage{footnote}}
\makesavenoteenv{longtable}
\usepackage{graphicx}
\makeatletter
\def\maxwidth{\ifdim\Gin@nat@width>\linewidth\linewidth\else\Gin@nat@width\fi}
\def\maxheight{\ifdim\Gin@nat@height>\textheight\textheight\else\Gin@nat@height\fi}
\makeatother
\setkeys{Gin}{width=\maxwidth,height=\maxheight,keepaspectratio}
\makeatletter
\def\fps@figure{htbp}
\makeatother
\NewDocumentCommand\citeproctext{}{}

\makeatletter
 \let\@cite@ofmt\@firstofone
 \def\@biblabel#1{}
 \def\@cite#1#2{{#1\if@tempswa , #2\fi}}
\makeatother
\newlength{\cslhangindent}
\newlength{\csllabelwidth}
\newenvironment{CSLReferences}[2] 
 {\begin{list}{}{%
  \setlength{\itemindent}{0pt}
  \setlength{\leftmargin}{0pt}
  \setlength{\parsep}{0pt}
  \ifodd #1
   \setlength{\leftmargin}{\cslhangindent}
   \setlength{\itemindent}{-1\cslhangindent}
  \fi
  \setlength{\itemsep}{#2\baselineskip}}}
 {\end{list}}
\usepackage{calc}

\usepackage{tikz}
\usepackage{xcolor}
\usepackage{tabularx}
\usepackage{orcidlink}
\usepackage{lineno}
\usepackage{libertinus}
\makeatletter
\@ifpackageloaded{caption}{}{\usepackage{caption}}
\AtBeginDocument{%
\ifdefined\contentsname
  \renewcommand*\contentsname{Table of contents}
\else
  \newcommand\contentsname{Table of contents}
\fi
\ifdefined\listfigurename
  \renewcommand*\listfigurename{List of Figures}
\else
  \newcommand\listfigurename{List of Figures}
\fi
\ifdefined\listtablename
  \renewcommand*\listtablename{List of Tables}
\else
  \newcommand\listtablename{List of Tables}
\fi
\ifdefined\figurename
  \renewcommand*\figurename{Figure}
\else
  \newcommand\figurename{Figure}
\fi
\ifdefined\tablename
  \renewcommand*\tablename{Table}
\else
  \newcommand\tablename{Table}
\fi
}
\@ifpackageloaded{float}{}{\usepackage{float}}
\floatstyle{ruled}
\@ifundefined{c@chapter}{\newfloat{codelisting}{h}{lop}}{\newfloat{codelisting}{h}{lop}[chapter]}
\floatname{codelisting}{Listing}

\makeatother
\makeatletter
\@ifpackageloaded{caption}{}{\usepackage{caption}}
\@ifpackageloaded{subcaption}{}{\usepackage{subcaption}}
\makeatother
\makeatletter
\@ifpackageloaded{algorithm}{}{\usepackage{algorithm}}
\makeatother
\makeatletter
\@ifpackageloaded{algpseudocode}{}{\usepackage{algpseudocode}}
\makeatother
\ifLuaTeX
  \usepackage{selnolig}  
\fi
\usepackage{bookmark}

\IfFileExists{xurl.sty}{\usepackage{xurl}}{} 
\hypersetup{
  pdftitle={Spectral Bridges},
  pdfauthor={Félix Laplante; Christophe Ambroise},
  pdfkeywords={spectral clustering, vector
quantization, scalable, non-parametric},
  colorlinks=true,
  linkcolor={blue},
  filecolor={Maroon},
  citecolor={Blue},
  urlcolor={Blue},
  pdfcreator={LaTeX via pandoc}}

\title{Spectral Bridges}
\usepackage{etoolbox}
\makeatletter
\providecommand{\subtitle}[1]{
  \apptocmd{\@title}{\par {\large #1 \par}}{}{}
}
\makeatother
\subtitle{Scalable Spectral Clustering free from hyperparameters}
\author{Félix Laplante \and Christophe Ambroise}

\begin{document}
\definecolor{computo-blue}{HTML}{034E79}


\title{Spectral Bridges}
\maketitle

\begin{center}
          Félix Laplante\quad
             Université de Paris Saclay\\
                 Christophe
Ambroise~\orcidlink{0000-0002-8148-0346}\footnote{Corresponding author: \href{mailto:christophe.ambroise@univ-evry.fr}{christophe.ambroise@univ-evry.fr}}\quad
             Laboratoire de Mathématiques et Modélisation
d'Evry, Université Paris-Saclay, CNRS, Univ Evry,\\
           
  \bigskip
  
  Date published: 2025-07-06 \quad Last modified: 2024-07-10
\end{center}
      
\bigskip
\begin{abstract}
In this paper, Spectral Bridges, a novel clustering algorithm, is
introduced. This algorithm builds upon the traditional k-means and
spectral clustering frameworks by subdividing data into small Voronoï
regions, which are subsequently merged according to a connectivity
measure. Drawing inspiration from Support Vector Machine's margin
concept, a non-parametric clustering approach is proposed, building an
affinity margin between each pair of Voronoï regions. This approach is
characterized by minimal hyperparameters and delineation of intricate,
non-convex cluster structures.

The numerical experiments underscore Spectral Bridges as a fast, robust,
and versatile tool for sophisticated clustering tasks spanning diverse
domains. Its efficacy extends to large-scale scenarios encompassing both
real-world and synthetic datasets.

The Spectral Bridge algorithm is implemented both in Python
(\url{https://pypi.org/project/spectral-bridges}) and R
\url{https://github.com/cambroise/spectral-bridges-Rpackage}).
\end{abstract}

\noindent%
{\it Keywords:} spectral clustering, vector
quantization, scalable, non-parametric
\vfill


\floatname{algorithm}{Algorithm}

\renewcommand*\contentsname{Contents}
{
\hypersetup{linkcolor=}
\setcounter{tocdepth}{3}
\tableofcontents
}
\section{Introduction}\label{introduction}

Clustering is a fundamental technique for exploratory data analysis,
organizing a set of objects into distinct homogeneous groups known as
clusters. It is extensively utilized across various fields, such as
biology for gene expression analysis (Eisen et al. 1998), social
sciences for community detection in social networks (Latouche, Birmelé,
and Ambroise 2011), and psychology for identifying behavioral patterns.
Clustering is often employed alongside supervised learning as a
pre-processing step, helping to structure and simplify data, thus
enhancing the performance and interpretability of subsequent predictive
models (Verhaak et al. 2010). Additionally, clustering can be integrated
into supervised learning algorithms, such as mixture of experts (Jacobs
et al. 1991), as part of a multi-objective strategy.

There are various approaches to clustering, and the quality of the
results is largely determined by how the similarity between objects is
defined, either through a similarity measure or a distance metric.
Clustering techniques originate from diverse fields of research, such as
genetics, psychometry, statistics, and computer science. Some methods
are entirely heuristic, while others aim to optimize specific criteria
and can be related to statistical models.

Density-based methods identify regions within the data with a high
concentration of points, corresponding to the modes of the joint
density. A notable non-parametric example of this approach is DBSCAN
(Ester et al. 1996). In contrast, model-based clustering, such as
Gaussian mixture models, represents a parametric approach to
density-based methods. Model-based clustering assumes that the data is
generated from a mixture of underlying probability distributions,
typically Gaussian distributions. Each cluster is viewed as a component
of this mixture model, and the Expectation-Maximization (EM) algorithm
is often used to estimate the parameters. This approach provides a
probabilistic framework for clustering, allowing for the incorporation
of prior knowledge and the ability to handle more complex cluster shapes
and distributions (McLachlan and Peel 2000).

Geometric approaches, such as k-means (MacQueen et al. 1967), are
distance-based methods that aim to partition data by optimizing a
criterion reflecting group homogeneity. The k-means++ algorithm (Arthur
and Vassilvitskii 2006) enhances this approach by providing faster and
more reliable results. However, a key limitation of these methods is the
assumption of linear boundaries between clusters, implying that clusters
are convex. To address non-convex clusters, the kernel trick can be
applied, allowing for a more flexible k-means algorithm. This approach
is comparable to spectral clustering in handling complex cluster
boundaries (Dhillon, Guan, and Kulis 2004). The k-means algorithm can
also be interpreted within the framework of model-based clustering under
specific assumptions (Govaert and Nadif 2003), revealing that it is
essentially a special case of the more general Gaussian mixture models,
where clusters are assumed to be spherical Gaussian distributions with
equal variance.

Graph-based methods represent data as a graph, with vertices symbolizing
data points and edges weighted to indicate the affinity between these
points. Spectral clustering can be seen as a relaxed version of the
graph cut algorithm (Shi and Malik 2000). However, traditional spectral
clustering faces significant limitations due to its high time and space
complexity, greatly hindering its applicability to large-scale problems
(Von Luxburg 2007).

The method we propose aims to find non-convex clusters in large
datasets, without relying on a parametric model, by using spectral
clustering based on an affinity that characterizes the local density of
the data. The algorithm described in this paper draws from numerous
clustering approaches. The initial intuition is to detect high-density
areas. To this end, vector quantization is used to divide the space into
a Voronoï tessellation. An original geometric criterion is then employed
to detect pairs of Voronoï regions that are either distant from each
other or separated by a low-density boundary. Finally, this affinity
measure is considered as the weight of an edge in a complete graph
connecting the centroids of the tessellation, and a spectral clustering
algorithm is used to find a partition of this graph. The only parameters
of the algorithm are the number of Voronoï Cells and the number of
clusters.

The paper begins with a section dedicated to presenting the context and
related algorithms, followed by a detailed description of the proposed
algorithm. Experiments and comparisons with reference algorithms are
then conducted on both real and synthetic data.

\section{Related Work}\label{related-work}

Spectral clustering is a graph-based approach that computes the
eigen-vectors of the graph's Laplacian matrix. This technique transforms
the data into a lower-dimensional space, making the clusters more
discernible. A standard algorithm like k-means is then applied to these
transformed features to identify the clusters (Von Luxburg 2007).
Spectral clustering enables capturing complex data structures and
discerning clusters based on the connectivity of data points in a
transformed space, effectively treating it as a relaxed graph cut
problem.

Classical spectral clustering involves two phases: construction of the
affinity matrix and eigen-decomposition. Constructing the affinity
matrix requires \(O(n^2d)\) time and \(O(n^2)\) memory, while
eigen-decomposition demands \(O(n^3)\) time and \(O(n^2)\) memory, where
\(n\) is the data size and \(d\) is the dimension. As \(n\) increases,
the computational load escalates significantly (Von Luxburg 2007).

To mitigate this computational burden, one common approach is to
sparsify the affinity matrix and use sparse eigen-solvers, reducing
memory costs but still requiring computation of all original matrix
entries (Von Luxburg 2007). Another strategy is sub-matrix construction.
The Nyström method randomly selects \(m\) representatives from the
dataset to form an \(n\times m\) affinity sub-matrix (Chen et al. 2010).
Cai et al.~extended this with the landmark-based spectral clustering
method, which uses k-means to determine \(m\) cluster centers as
representatives (Cai and Chen 2014). Ultra-scalable spectral clustering
(U-SPEC) employs a hybrid representative selection strategy and a fast
approximation method for constructing a sparse affinity sub-matrix
(Huang et al. 2019).

Other approaches use the properties of the small initial clusters for
the affinity computation. Clustering Based on Graph of Intensity
Topology (GIT) estimates for example a global topological graph
(topo-graph) between local clusters (Gao et al. 2021). It then uses the
Wasserstein Distance between predicted and prior class proportions to
automatically cut noisy edges in the topo-graph and merge connected
local clusters into final clusters.

The issue of characterizing the affinity between two clusters to create
an edge weight is central to the efficiency of a spectral clustering
algorithm operating from a submatrix.

Notice that the clustering robustness of many Spectral clustering
algorithms heavily relies on the proper selection of kernel parameter,
which is difficult to find without prior knowledge (Ng, Jordan, and
Weiss 2001).

\section{Spectral Bridges}\label{spectral-bridges}

The proposed algorithm uses k-means centroids for vector quantization
defining Voronoï region, and a strategy is proposed to link these
regions, with an ``affinity'' gauged in terms of minimal margin between
pairs of classes. These affinities are considered as weight of edges
defining a completely connected graph whose vertices are the regions.
Spectral clustering on the region provide a partition of the input
space. The sole parameters of the algorithm are the number of Voronoï
region and the number of final cluster.

\subsection{Bridge affinity}\label{sec-bridge-affinity}

\begin{figure}

\begin{minipage}{0.50\linewidth}
\includegraphics{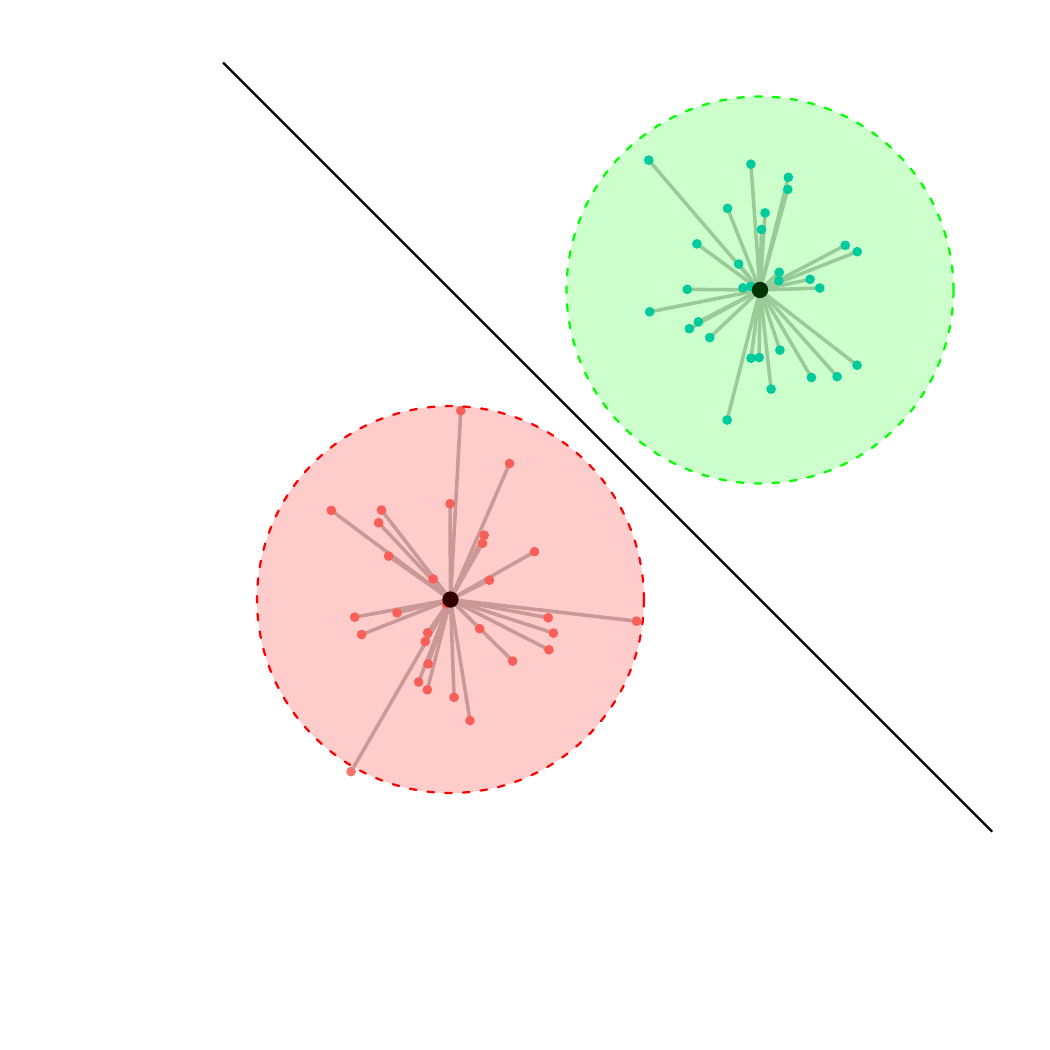}\end{minipage}%
\begin{minipage}{0.50\linewidth}
\includegraphics{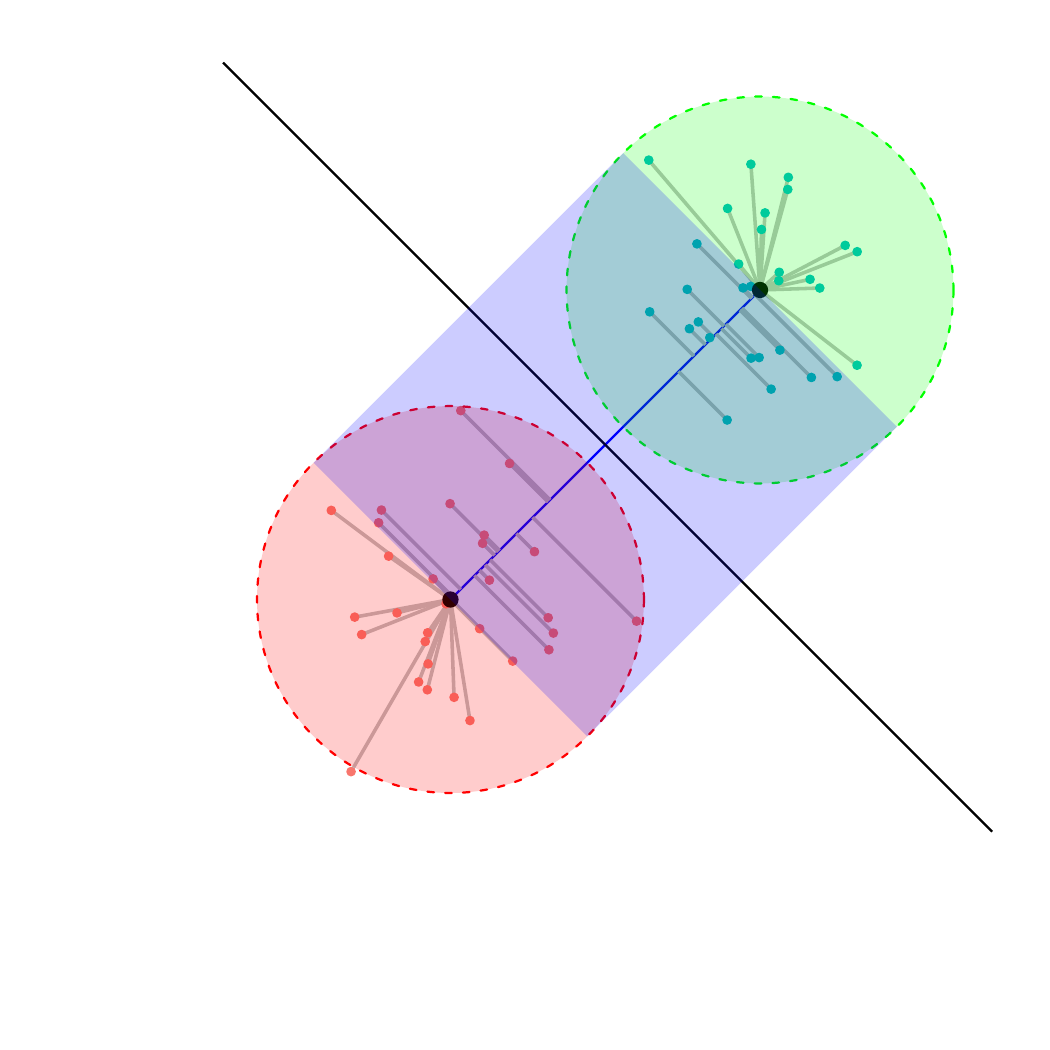}\end{minipage}%

\caption{\label{fig-balls-bridge}Balls (left) versus Bridge (right). The
inertia of each structure is the sum of the squared distances
represented by grey lines.}

\end{figure}

The basic idea involves calculating the difference in inertia achieved
by projecting onto a segment connecting two centroids, rather than using
the two centroids separately (see Figure~\ref{fig-balls-bridge}). If the
difference is small, it suggests a low density between the classes.
Conversely, if this diffrence is large, it indicates that the two
classes may reside within the same densely populated region.

Let us consider a sample \(X=(\boldsymbol x_i)_{i \in \{1,\cdots,n\}}\)
of vectors \(\boldsymbol x_i \in \mathbb R^d\) and a set of \(m\) coding
vectors \((\boldsymbol \mu_k)_{k \in \{1,\cdots,m\}}\) defining a
partition \(P=\{\mathcal{V}_1,\cdots,\mathcal{V}_m \}\) of
\(\mathbb R^d\) into \(m\) Voronoï regions: \[
\mathcal{V}_k = \left\{ \mathbf{x} \in \mathbb{R}^d \mid \|\mathbf{x} - \boldsymbol{\mu}_k\| \leq \|\mathbf{x} - \boldsymbol{\mu}_j\| \text{ for all } j \neq k \right\}.
\]

In the following a ball denotes the subset of \(X\) in a Voronoï region.
The inertia of two balls \(\mathcal{V}_k\) and \(\mathcal{V}_l\) is \[
I_{kl} = \sum_{\boldsymbol x_i\in \mathcal{V}_k} \|\boldsymbol x_i - \boldsymbol \mu_k\|^2  + \sum_{\boldsymbol x_i\in \mathcal{V}_l} \|\boldsymbol x_i - \boldsymbol \mu_l\|^2.
\] We define a bridge as a structure defined by a segment connecting two
centroids \(\boldsymbol \mu_k\) and \(\boldsymbol \mu_l\). The inertia
of a bridge between \(\mathcal{V}_k\) and \(\mathcal{V}_l\) is defined
as \[
B_{kl} = \sum_{\boldsymbol x_i\in \mathcal{V}_k \cup \mathcal{V}_l} \|\boldsymbol x_i - \boldsymbol p_{kl}(\boldsymbol x_i)\|^2,
\] where \[
\boldsymbol p_{kl}(\boldsymbol x_i) = \boldsymbol \mu_{k} + t_i(\boldsymbol \mu_{l} - \boldsymbol \mu_{k}),
\] with \[
t_i  = \min\left(1, \max\left(0, \frac{\langle \boldsymbol x_i - \boldsymbol \mu_k | \boldsymbol \mu_l - \boldsymbol \mu_k\rangle}{\|  \boldsymbol \mu_l - \boldsymbol \mu_k \|^2}\right)\right). 
\]

Considering two centroïds, the normalized average of the difference
betweenn Bridge and balls inertia (See \hyperref[gain]{Appendix})
constitutes the basis of our affinity measure between two regions: \[
\begin{aligned}
\frac{B_{kl}- I_{kl}}{(n_k+n_l)\|\boldsymbol \mu_k - \boldsymbol \mu_l\|^2} &=& \frac{\sum_{\boldsymbol{x_i} \in \mathcal V_k} \langle \boldsymbol{x_i} - \boldsymbol{\mu}_k \vert \boldsymbol{\mu}_l - \boldsymbol{\mu}_k \rangle_+^2  \sum_{\boldsymbol{x_i} \in \mathcal V_l} \langle \boldsymbol{x_i} - \boldsymbol{\mu}_l \vert \boldsymbol{\mu}_k - \boldsymbol{\mu}_l\rangle_+^2}{(n_k+n_l)\|\boldsymbol \mu_k - \boldsymbol \mu_l\|^4},\\
&=&  \frac{\sum_{\boldsymbol{x_i} \in \mathcal V_k \cup \mathcal V_l} \alpha_i^2}{n_k+n_l},
\end{aligned}
\] where \[
\alpha_i=
\begin{cases}
t_i, &  \text{ if } t_i\in[0,1/2],\\
1-t_i, & \text{ if } t_i\in]1/2,1].
\end{cases}
\]

The basic intuition behind this affinity is that \(t_i\) represents the
relative position of the projection of \(\boldsymbol x_i\) on the
segment \([\boldsymbol \mu_k,\boldsymbol \mu_l]\). \(\alpha_i\)
represents the relative position on the segment, with the centroid of
the class to which \(\boldsymbol x_i\) belongs as the reference point.

The boundary that separates the two clusters defined by centroids
\(\boldsymbol \mu_k\) and \(\boldsymbol \mu_l\) is a hyperplane. This
hyperplane is orthogonal to the line segment connecting the centroids
and intersects this segment at its midpoint.

If we consider all points
\(\boldsymbol x_i \in \mathcal V_k \cup \mathcal V_l\) which are not
projected on centroids but somewhere on the segment, the distance from a
point to the hyperplane is \[
\|\boldsymbol p_{kl}(\boldsymbol x_i) - \boldsymbol \mu_{kl}\| = (1/2-\alpha_i) \| \boldsymbol \mu_k-\boldsymbol \mu_l \|.
\]

This distance is similar to the concept of margin in Support Vector
Machine (Cortes and Vapnik 1995). When the \(\alpha_i\) values are small
(close to zero since \(\alpha_i\in [0,1/2]\)), the margins to the
hyperplane are large, indicating a low density between the classes.
Conversely, if the margins are small, it suggests that the two classes
may reside within the same densely populated region. Consequently, the
sum of the \(\alpha_i\) or \(\alpha_i^2\) increases with the density of
the region between the classes.

Note that the criterion is local and indicates the relative difference
in densities between the balls and the bridge, rather than evaluating a
global score for the densities of the structures.

Eventually, we define the bridge affinity between centroids \(k\) and
\(l\) as: \[
a_{kl}=
\begin{cases}
0, & \text{ if } k=l,\\
 \frac{\sum_{\boldsymbol{x_i} \in \mathcal V_k \cup \mathcal V_l} \alpha_i^2}{n_k+n_l}, & \text{otherwise}.
\end{cases}
\] To allow points with large margin to dominate and make the algorithm
more robust to noise and outliers we consider the following exponential
transformation: \[
\tilde{a}_{kl} = g(a_{kl})=\exp(\gamma\sqrt{a_{kl}}).
\]

where \(\gamma\) is a scaling factor. This factor is set to ensure a
large enough separation between the final coefficients. This factor is
determined by the equation:
\[
\gamma = \frac{\log(M)}{\sqrt{q_{90}} - \sqrt{q_{10}}}
\]

where \(q_{10}\) and \(q_{90}\) are respectively the 10th and 90th
percentiles of the original affinity matrix and \(M > 1\). Thus, since
the transformation is order-preserving, the 90th percentile of the newly
constructed matrix is \(M\) times greater than the 10th percentile. By
default, \(M\) is arbitrarily set to a large value of \(10^4\).

The inclusion of the square root can be understood as redefining the
affinity measure. Instead of considering the variance and the squared
Euclidean norm, we interpret the affinity as the ratio between the
standard deviation and the length of the segment connecting two
centroids. This reinterpretation greatly enhances numerical stability,
contributing to more reliable clustering results.

\subsection{Algorithm}\label{algorithm}

\begin{figure}

\begin{minipage}{0.33\linewidth}

\includegraphics{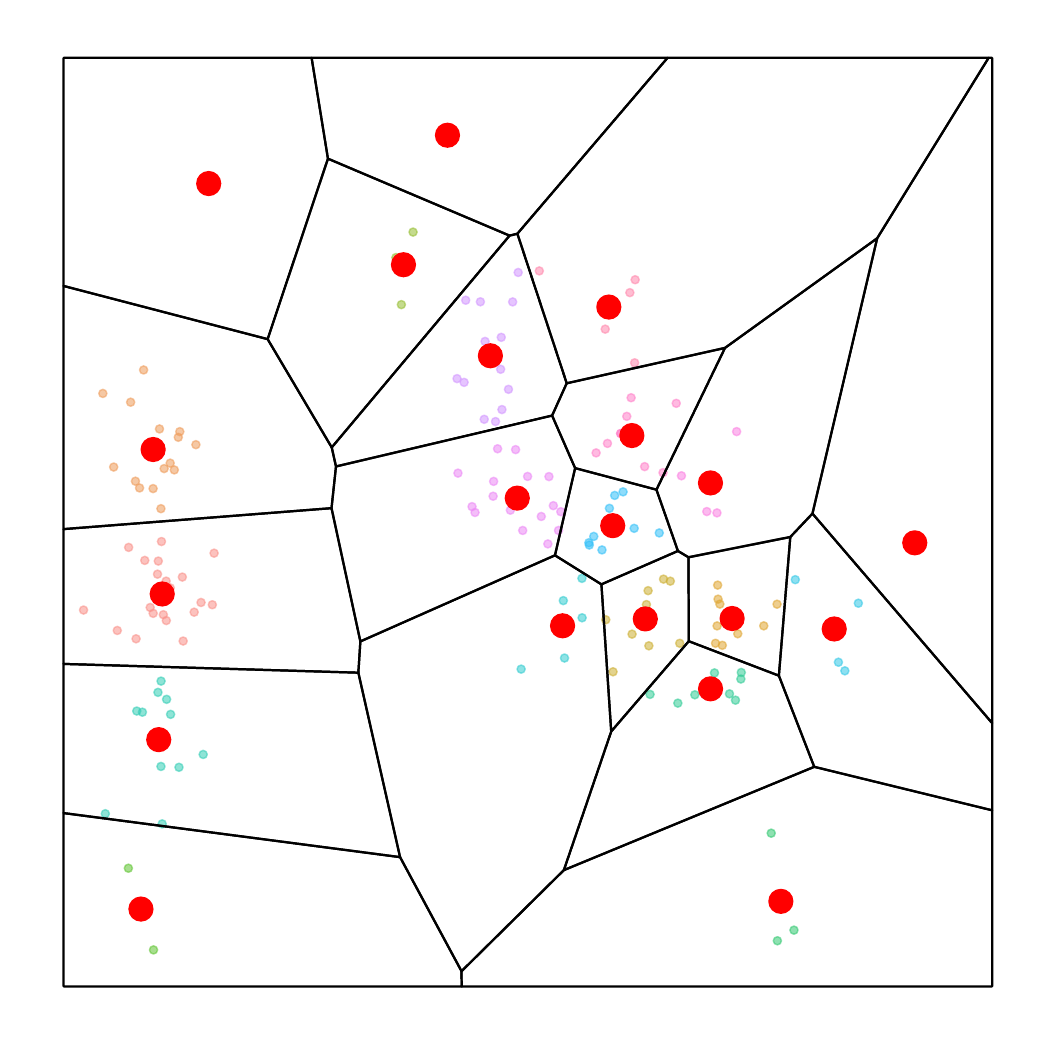}

\subcaption{\label{}Vector quantization}
\end{minipage}%
\begin{minipage}{0.33\linewidth}

\includegraphics{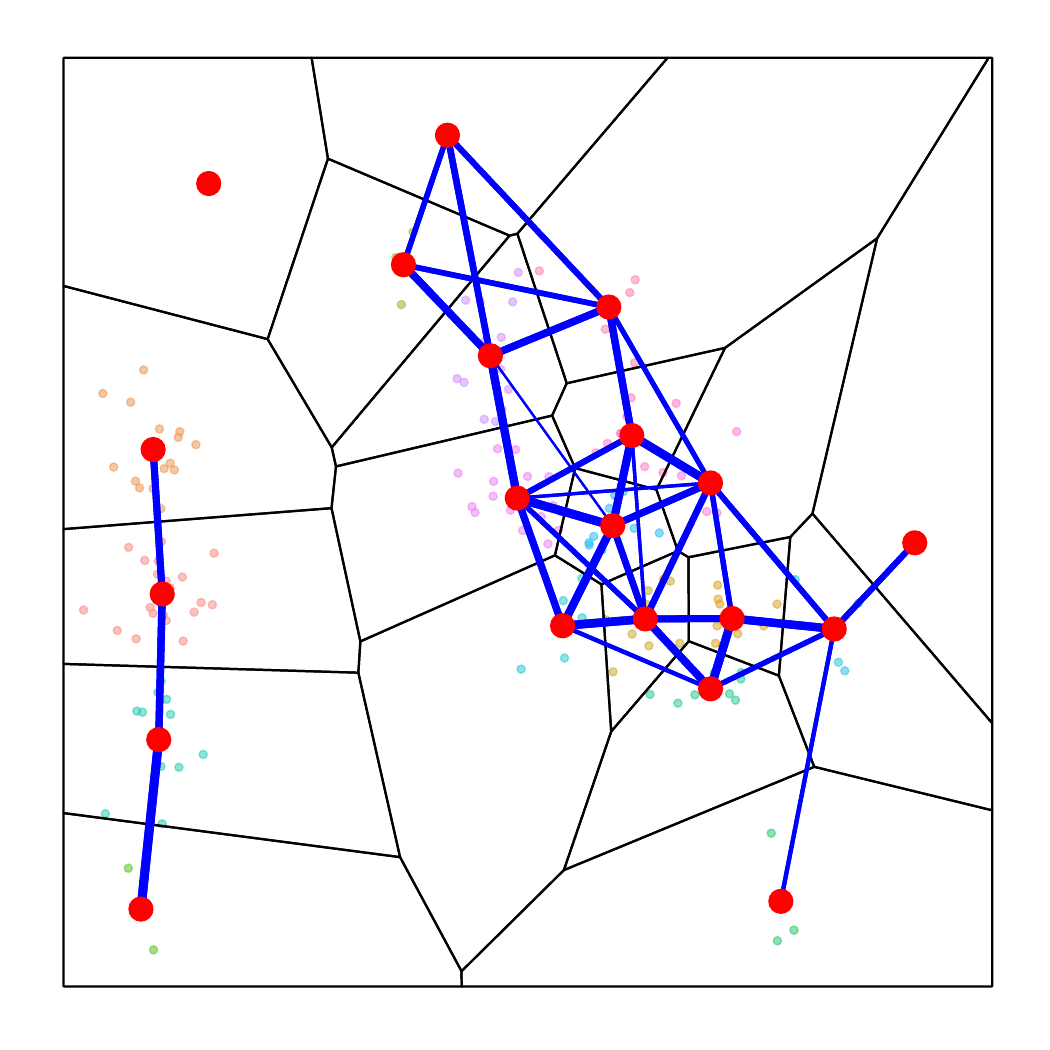}

\subcaption{\label{}Affinity computation}
\end{minipage}%
\begin{minipage}{0.33\linewidth}

\includegraphics{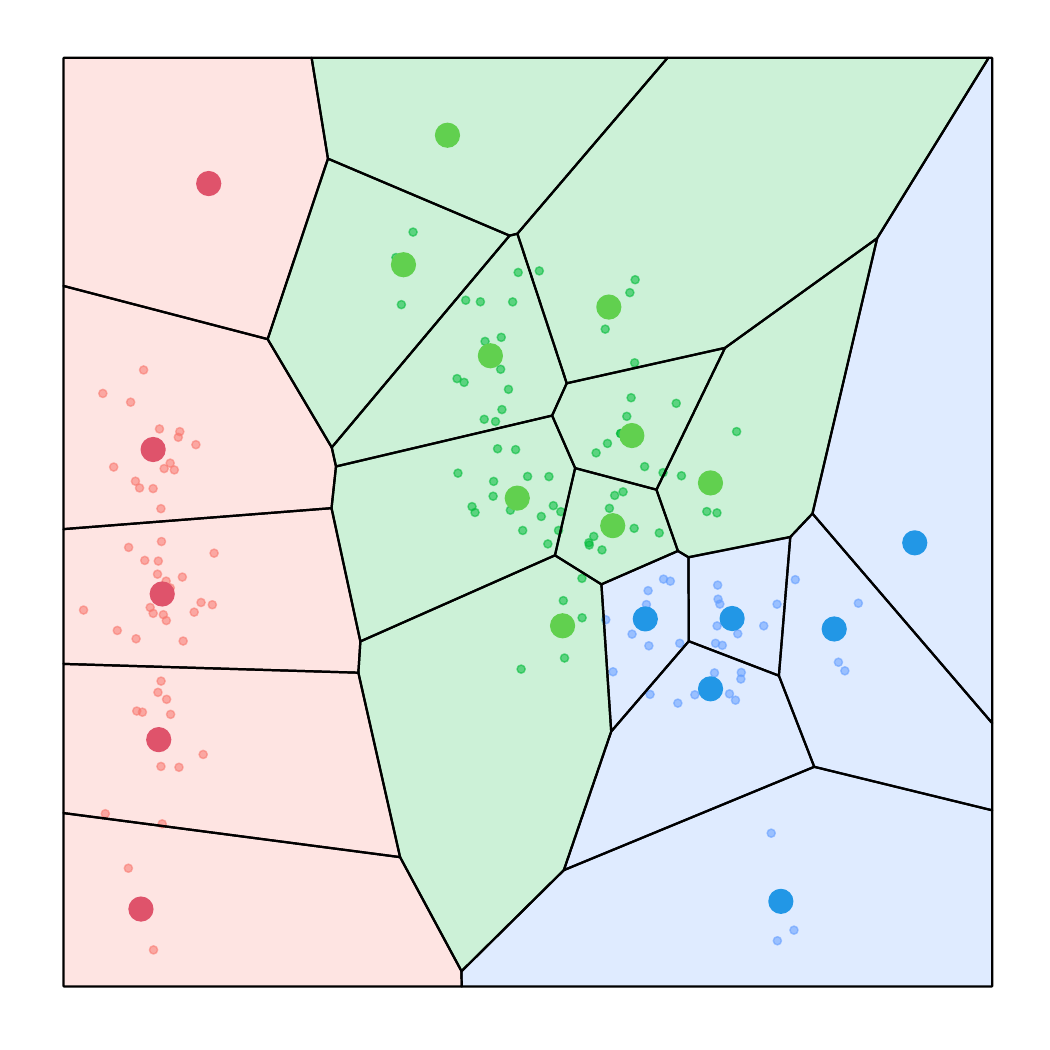}

\subcaption{\label{}Spectral clustering}
\end{minipage}%

\caption{\label{fig-steps}Illustration of the Spectral bridges algorithm
with the Iris dataset (first principal plane). Vector quantization (Step
1 of  Algorithm~\ref{alg-spectral-bridges} ), Affinity computation (Step
2 of  Algorithm~\ref{alg-spectral-bridges} ), Spectral clustering and
spreading (Step 3-4 of  Algorithm~\ref{alg-spectral-bridges} ).}

\end{figure}

The Spectral Bridges algorithm first identifies local clusters to define
Voronoï regions, computes edges with affinity weights between these
regions, and ultimately cuts edges between regions with low inter-region
density to determine the final clusters (See
 Algorithm~\ref{alg-spectral-bridges}  and Figure~\ref{fig-steps}).

In spectral clustering, the time complexity is usually dominated by the
eigen-decomposition step, which is \(O(n^3)\). However, in the case of
Spectral Bridges, the k-means algorithm has a time complexity of
\(O(n \times m \times d)\). For datasets with large \(n\), this can be
more significant than the \(O(m^3)\) time complexity of the Spectral
Bridges eigen-decomposition. As for the affinity matrix construction,
there are \(m^2\) coefficients to be calculated. Each \(a_{kl}\)
coefficient requires the computation of \(n_k + n_l\) dot products as
well as the norm \(\| \boldsymbol \mu_k-\boldsymbol \mu_l \|\), the
latter often being negligeable. Assuming that the Voronoï regions are
roughly balanced in cardinality, we have \(n_k \approx \frac{n}{m}\).
Since \(m\) should always be less than \(n\), therefore
\(\frac{n}{m} > 1\) and the time complexity of the affinity matrix is
\(O(\frac{n}{m} \times m^2 \times d) = O(n \times m \times d)\) given
the acceptable range of values for \(m\). Nonetheless, this is rarely
the bottleneck.

\begin{algorithm}[htb!]
\caption{Spectral Bridges}
\label{alg-spectral-bridges}
\begin{algorithmic}[1]
\Procedure{SpectralBridges}{$X, k, m$}
\Comment{$X$: input dataset, $k$: number of clusters, $m$: number of Voronoï regions}

    \State \textbf{Step 1: Vector Quantization}
    \State $\text{centroids}, \text{voronoiRegions} \gets$ \Call{KMeans}{$X, m$}
    \Comment{Initial centroids and Voronoi regions using k-means++}

    \State \textbf{Step 2: Affinity Computation}
    \State $A = \{g(a_{kl})\}_{kl} \gets$ \Call{Affinity}{$X, \text{centroids}, \text{voronoiRegions}$}
    \Comment{Compute affinity matrix $A$}

    \State \textbf{Step 3: Spectral Clustering}
    \Comment{Assign each region to a cluster}
    \State $\text{labels} \gets$ \Call{SpectralClustering}{$A, k$}

    \State \textbf{Step 4: Propagate}
    \Comment{Assign each data point to the cluster of its region}
    \State $\text{clusters} \gets$ \Call{Propagate}{$X, \text{labels}, \text{voronoiRegions}$}

    \State \Return $\text{clusters}$
    \Comment{Return cluster labels for data points in $X$}
\EndProcedure
\end{algorithmic}
\end{algorithm}

\section{Numerical experiments}\label{numerical-experiments}

In this section, the results obtained from testing the Spectral Bridges
algorithm on various datasets, both small and large scale, including
real-world and well-known synthetic datasets, are presented. These
experiments assess the accuracy, time and space complexity, ease of use,
robustness, and adaptability of our algorithm. We compare Spectral
Bridges (SB) against several state-of-the-art methods, including
k-means++ (KM) (MacQueen et al. 1967; Arthur and Vassilvitskii 2006),
Expectation-Maximization (EM) (Dempster, Laird, and Rubin 1977), Ward
Clustering (WC) (Ward Jr 1963), and DBSCAN (DB) (Ester et al. 1996).
This comparison establishes baselines across centroid-based clustering
algorithms, hierarchical methods, and density-based methods.

The algorithms are evaluated on both raw and PCA-processed data with
varying dimensionality. For synthetic datasets, Gaussian and/or uniform
noise is introduced to assess the robustness of the algorithm.

\subsection{Datasets}\label{datasets}

\subsubsection{Real-world data}\label{real-world-data}

\begin{itemize}
\tightlist
\item
  \textbf{MNIST}: A large dataset containing 60,000 handwritten digit
  images in ten balanced classes, commonly used for image processing
  benchmarks. Each image consists of \(28 \times 28 = 784\) pixels.
\item
  \textbf{UCI ML Breast Cancer Wisconsin}: A dataset featuring computed
  attributes from digitized images of fine needle aspirates (FNA) of
  breast masses, used to predict whether a tumor is malignant or benign.
\end{itemize}

\subsubsection{Synthetic data}\label{synthetic-data}

\begin{itemize}
\tightlist
\item
  \textbf{Impossible}: A synthetic dataset designed to challenge
  clustering algorithms with complex patterns.
\item
  \textbf{Moons}: A two-dimensional dataset with two interleaving
  half-circles.
\item
  \textbf{Circles}: A synthetic dataset of points arranged in two
  non-linearly separable circles.
\item
  \textbf{Smile}: A synthetic dataset with points arranged in the shape
  of a smiling face, used to test the separation of non-linearly
  separable data.
\end{itemize}

\subsubsection{Datasets Summary \& Class
Balance}\label{datasets-summary-class-balance}

\begin{longtable}[]{@{}
  >{\raggedright\arraybackslash}p{(\columnwidth - 8\tabcolsep) * \real{0.1591}}
  >{\raggedright\arraybackslash}p{(\columnwidth - 8\tabcolsep) * \real{0.1023}}
  >{\raggedright\arraybackslash}p{(\columnwidth - 8\tabcolsep) * \real{0.1364}}
  >{\raggedright\arraybackslash}p{(\columnwidth - 8\tabcolsep) * \real{0.1364}}
  >{\raggedright\arraybackslash}p{(\columnwidth - 8\tabcolsep) * \real{0.4659}}@{}}
\caption{Datasets Summary \& Class Balance}\tabularnewline
\toprule\noalign{}
\begin{minipage}[b]{\linewidth}\raggedright
\textbf{Dataset}
\end{minipage} & \begin{minipage}[b]{\linewidth}\raggedright
\textbf{\#Dims}
\end{minipage} & \begin{minipage}[b]{\linewidth}\raggedright
\textbf{\#Samples}
\end{minipage} & \begin{minipage}[b]{\linewidth}\raggedright
\textbf{\#Classes}
\end{minipage} & \begin{minipage}[b]{\linewidth}\raggedright
\textbf{Class Proportions}
\end{minipage} \\
\midrule\noalign{}
\endfirsthead
\toprule\noalign{}
\begin{minipage}[b]{\linewidth}\raggedright
\textbf{Dataset}
\end{minipage} & \begin{minipage}[b]{\linewidth}\raggedright
\textbf{\#Dims}
\end{minipage} & \begin{minipage}[b]{\linewidth}\raggedright
\textbf{\#Samples}
\end{minipage} & \begin{minipage}[b]{\linewidth}\raggedright
\textbf{\#Classes}
\end{minipage} & \begin{minipage}[b]{\linewidth}\raggedright
\textbf{Class Proportions}
\end{minipage} \\
\midrule\noalign{}
\endhead
\bottomrule\noalign{}
\endlastfoot
MNIST & 784 & 60000 & 10 & 9.9\%, 11.2\%, 9.9\%, 10.3\%, 9.7\%, 9\%,
9.9\%, 10.4\%, 9.7\%, 9.9\% \\
Breast Cancer & 30 & 569 & 2 & 37.3\%, 62.7\% \\
Impossible & 2 & 3594 & 7 & 24.8\%, 18.8\%, 11.3\%, 7.5\%, 12.5\%,
12.5\%, 12.5\% \\
Moons & 2 & 1000 & 2 & 50\%, 50\% \\
Circles & 2 & 1000 & 2 & 50\%, 50\% \\
Smile & 2 & 1000 & 4 & 25\%, 25\%, 25\%, 25\% \\
\end{longtable}

Class proportions are presented in ascending order starting from label
\(0\).

\subsection{Metrics}\label{metrics}

To evaluate the performance of the clustering algorithm, the Adjusted
Rand Index (ARI) (Halkidi, Batistakis, and Vazirgiannis 2002) and
Normalized Mutual Information (NMI) (Cover and Thomas 1991) are used.
ARI measures the similarity between two clustering results, ranging from
-0.5 to 1, with 1 indicating perfect agreement. NMI ranges from 0 to 1,
with higher values indicating better clustering quality. In some tests,
the variability of scores across multiple runs is also reported due to
the random initialization in k-means, though k-means++ generally
provides stable and reproducible results.

\subsection{Platform}\label{platform}

All experiments were conducted on an Archlinux machine with Linux 6.9.3
Kernel, 8GB of RAM, and an AMD Ryzen 3 7320U processor.

\subsection{Hyperparameter settings}\label{hyperparameter-settings}

The hyperparameters of the Spectral Bridges algorithm were based on the
size of each dataset, \(n\), and the number of clusters, \(K\). A larger
number of clusters typically suggests that a higher value for the number
of Voronoï regions is optimal. Conversely, using a high number of
Voronoï regions for a small dataset might result in nearly empty regions
that do not adequately represent any local structure.

A good yet not very precise way of setting the number of Voronoï regions
\(m\) is to observe the Within Cluster Sum of Squares (WCSS) or inertia
in a way akin to the elbow method. Since \(m\) should be set to a value
strictly greater than \(K\), we plot the WCSS for varying values of
\(m\), and find a value such that the WCSS-\(m\) relationship becomes
quasi-linear.

By adjusting \(m\) in this manner, we aim to balance the need for
detailed representation with the risk of overfitting, ensuring that each
Voronoï region meaningfully captures the underlying data distribution.
The sensitivity or lack thereof is illustrated later on by
Figure~\ref{fig-m-vs-score}.

For other algorithms, such as DBSCAN, labels were used to determine the
best hyperparameter values to compare our method against the ``best case
scenario'', thus putting the Spectral Bridges algorithm at a voluntary
disadvantage.

\subsection{Time complexity}\label{time-complexity}

\begin{figure}

\begin{minipage}{0.50\linewidth}

\includegraphics{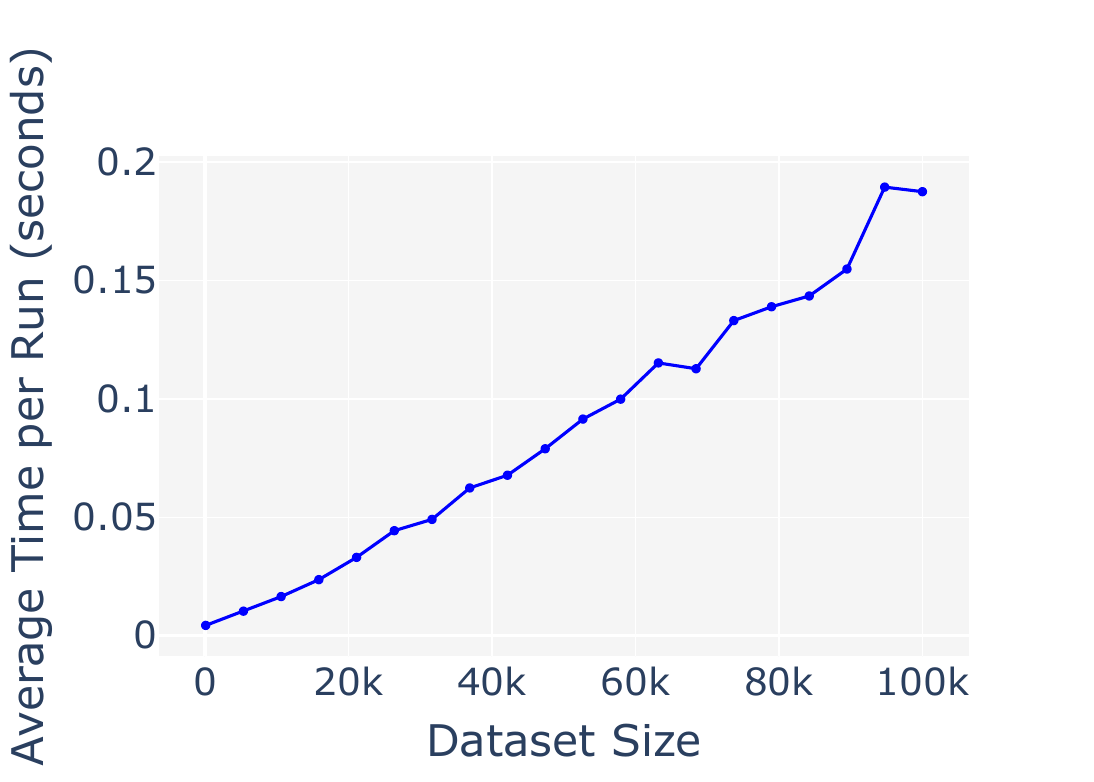}

\subcaption{\label{}Varying \(n\), fixed \(m = 10\)}
\end{minipage}%
\begin{minipage}{0.50\linewidth}

\includegraphics{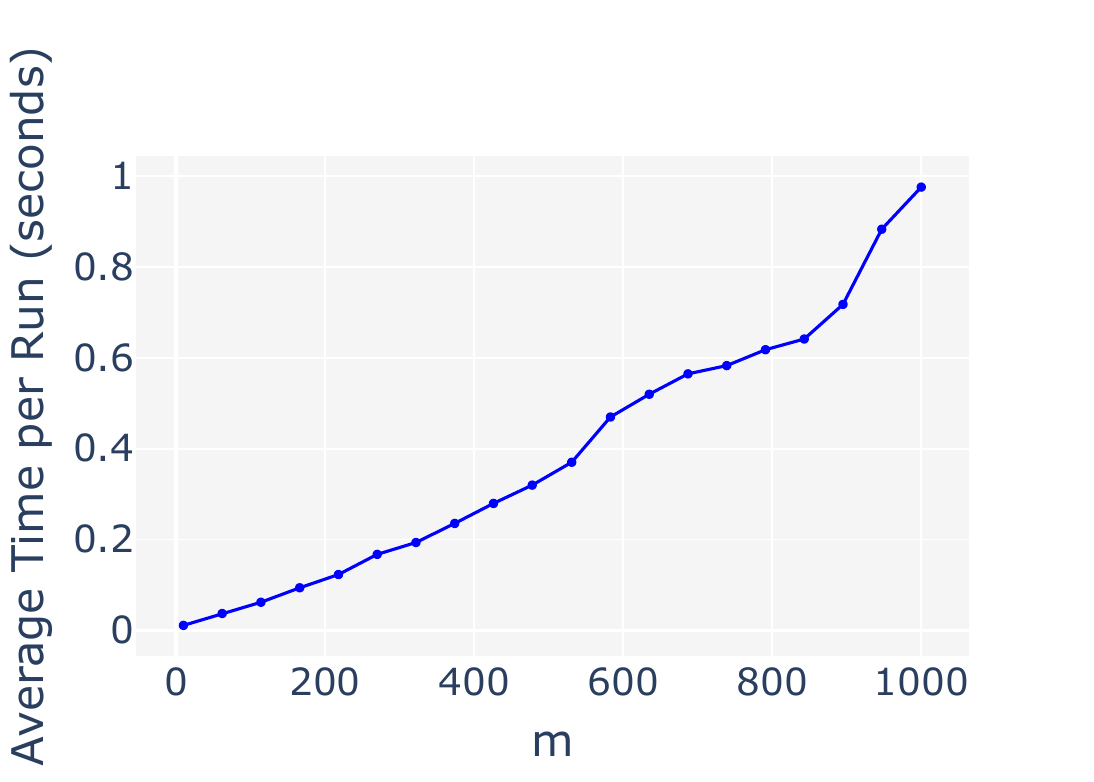}

\subcaption{\label{}Varying \(m\), fixed \(n = 5000\)}
\end{minipage}%

\caption{\label{fig-time-complexity}Average time taken per model fit.}

\end{figure}

To assess the algorithm's time complexity, the average execution times
over 50 runs were computed for varying numbers of Voronoï regions \(m\)
as well as dataset sizes. With a constant number of clusters \(K = 5\)
and an embedding dimension of \(d = 10\), the results (see
Figure~\ref{fig-time-complexity}) highlight Spectral Bridges algorihtm's
efficacy. As discussed previously, we observe a linear relationship
between \(m\) and the execution time because the matrix construction is
highly optimized and the time taken is almost negligeable compared to
that of the initial k-means++ centroids initalization.

\subsection{Accuracy}\label{accuracy}

The algorithm's accuracy was first evaluated on the MNIST dataset.
Metrics were collected to compare our method with k-means++, EM, and
Ward clustering. Metric were estimated by taking the empirical average
over 10 consecutive runs with the same random seed for each method.
Since our computational capabilites were too limited, a sample of 20,000
(one third) data points was chosen at random for each iteration.

\begin{figure}

\centering{

\includegraphics{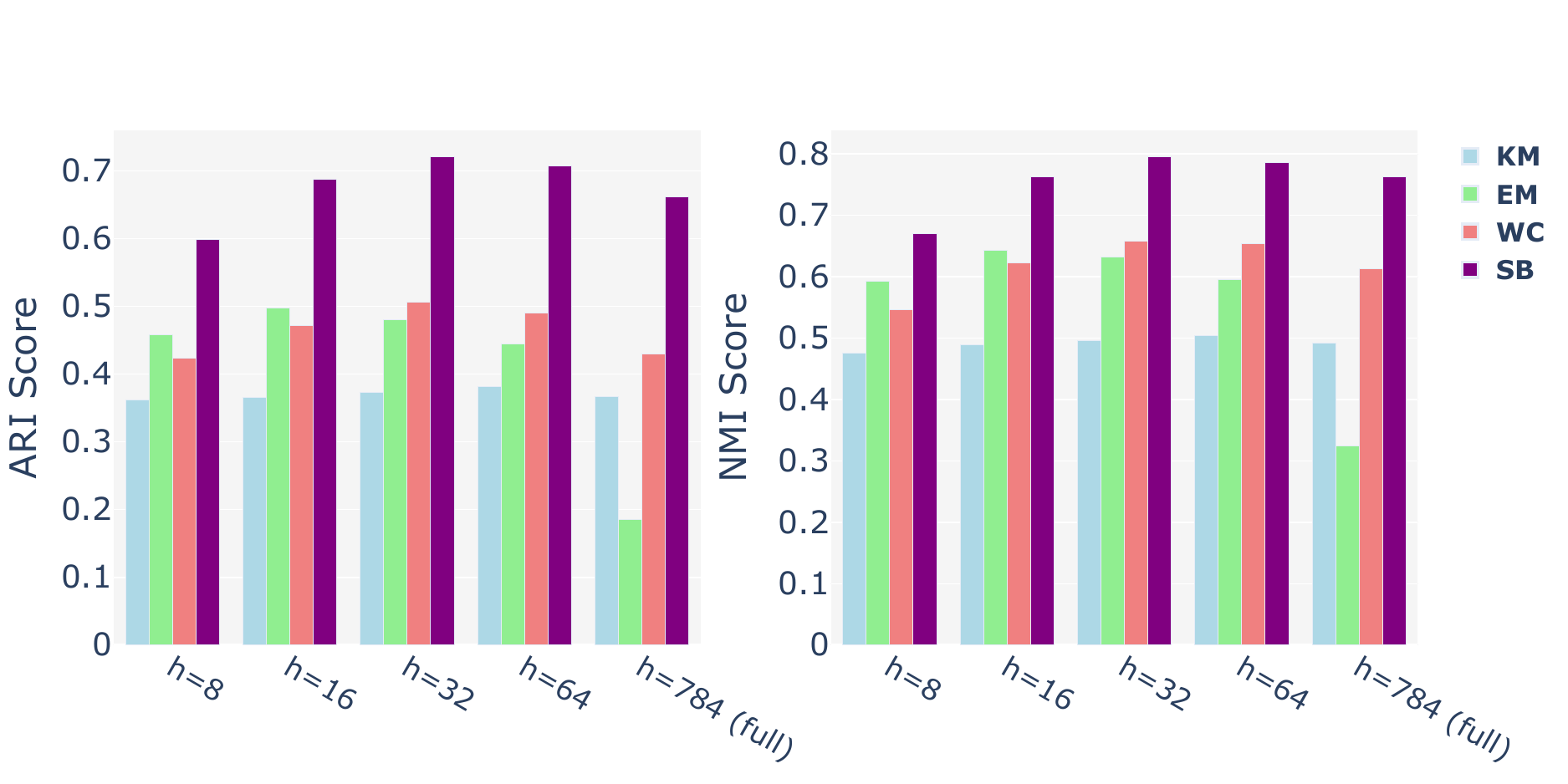}

}

\caption{\label{fig-mnist-scores}ARI and NMI scores of k-means++ (blue),
EM (green), Ward Clustering (red), and Spectral Bridges (purple) on PCA
embedding and full MNIST.}

\end{figure}

Let \(h\) denote the embedding dimension of the dataset. Spectral
Bridges was tested both on the raw MNIST dataset without preprocessing
(\(h = 784\)) and after reducing its dimension using PCA to
\(h \in \{8, 16, 32, 64\}\) (see Figure~\ref{fig-mnist-scores}).

\begin{figure}

\begin{minipage}{0.33\linewidth}

\includegraphics{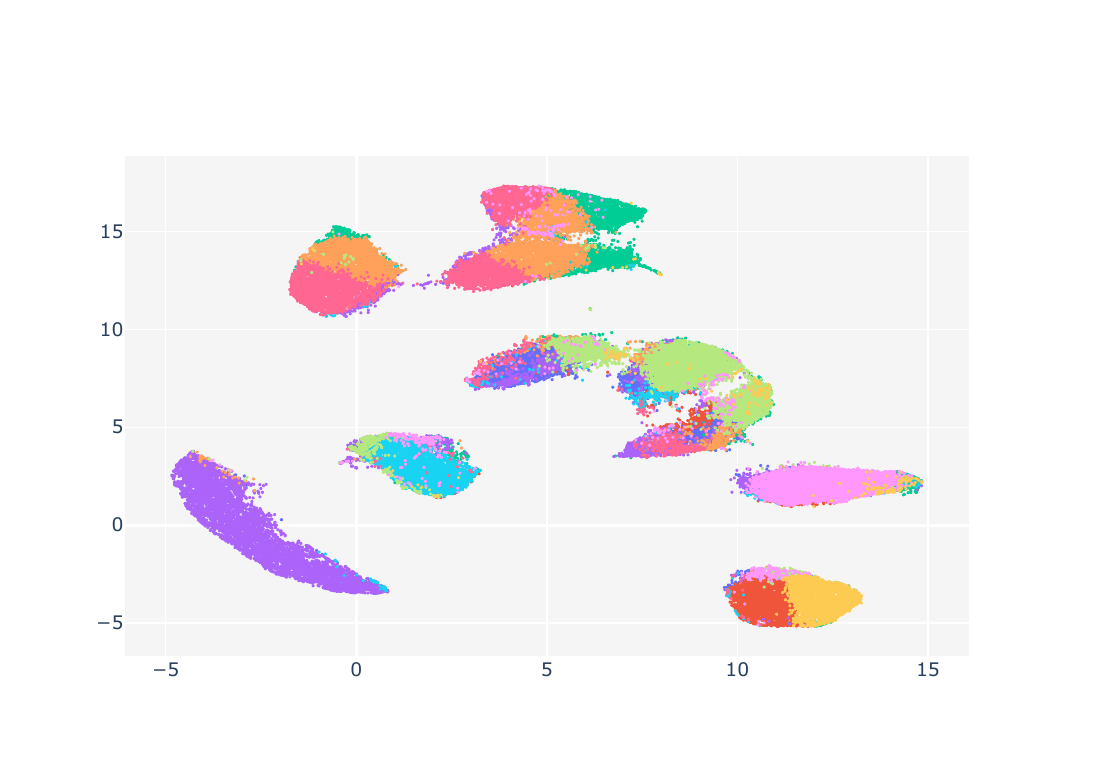}

\subcaption{\label{}k-means++}
\end{minipage}%
\begin{minipage}{0.33\linewidth}

\includegraphics{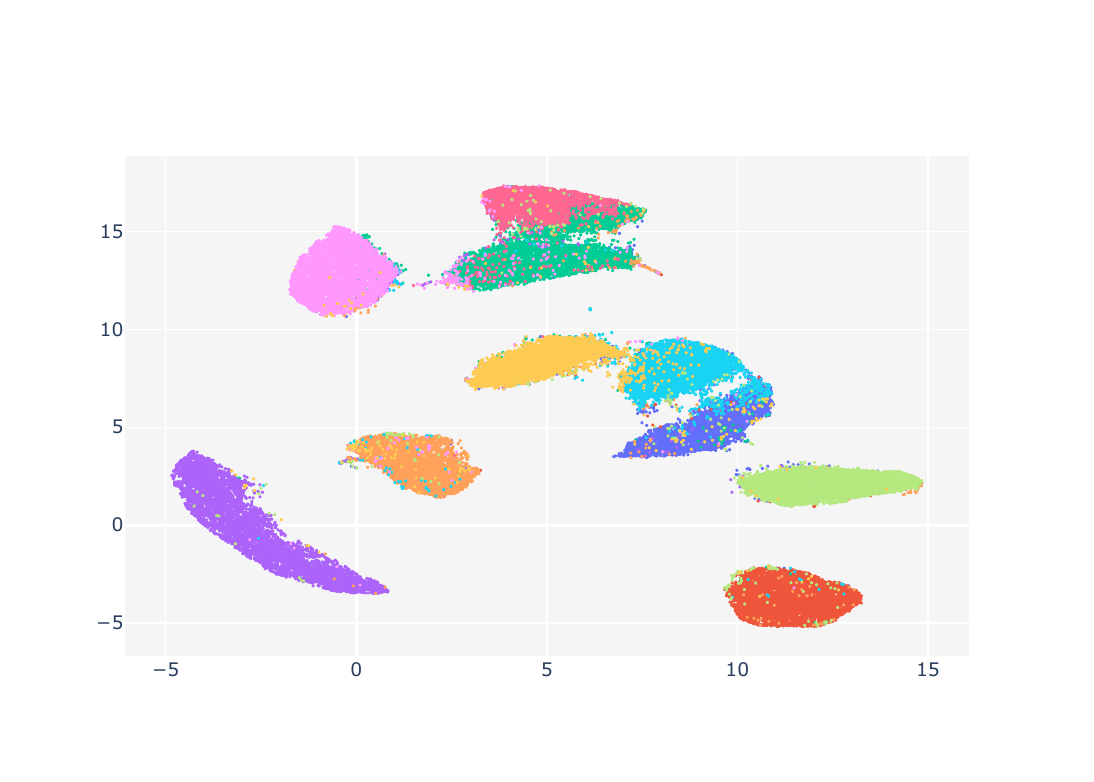}

\subcaption{\label{}Spectral Bridges}
\end{minipage}%
\begin{minipage}{0.33\linewidth}

\includegraphics{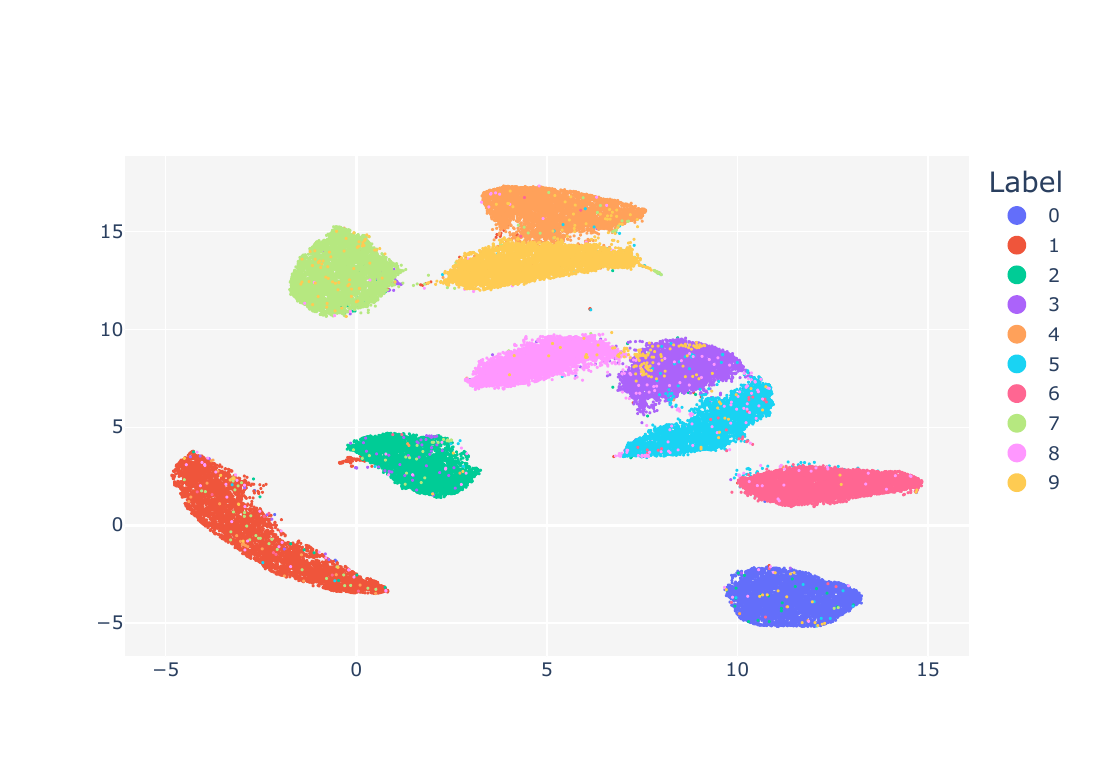}

\subcaption{\label{}Ground Truth}
\end{minipage}%

\caption{\label{fig-MNIST}UMAP projection of predicted clusters against
the ground truth labels.}

\end{figure}

For visualization purposes, the predicted clusters by Spectral Bridges
and k-means++ were projected using UMAP to compare them against the
ground truth labels and to better understand the cluster shapes (see
Figure~\ref{fig-MNIST}). Note that the projection was not used in the
experiments as an embedding, and thus does not play any role in the
clustering process itself. As a matter of fact, the embedding used was
obtained with PCA, \(h = 32\) and 250 Voronoï regions. Note that the
label colors match the legend only in the case of the ground truth data.
Indeed, the ordering of the labels have no significance on clustering
quality.

\begin{figure}

\centering{

\includegraphics{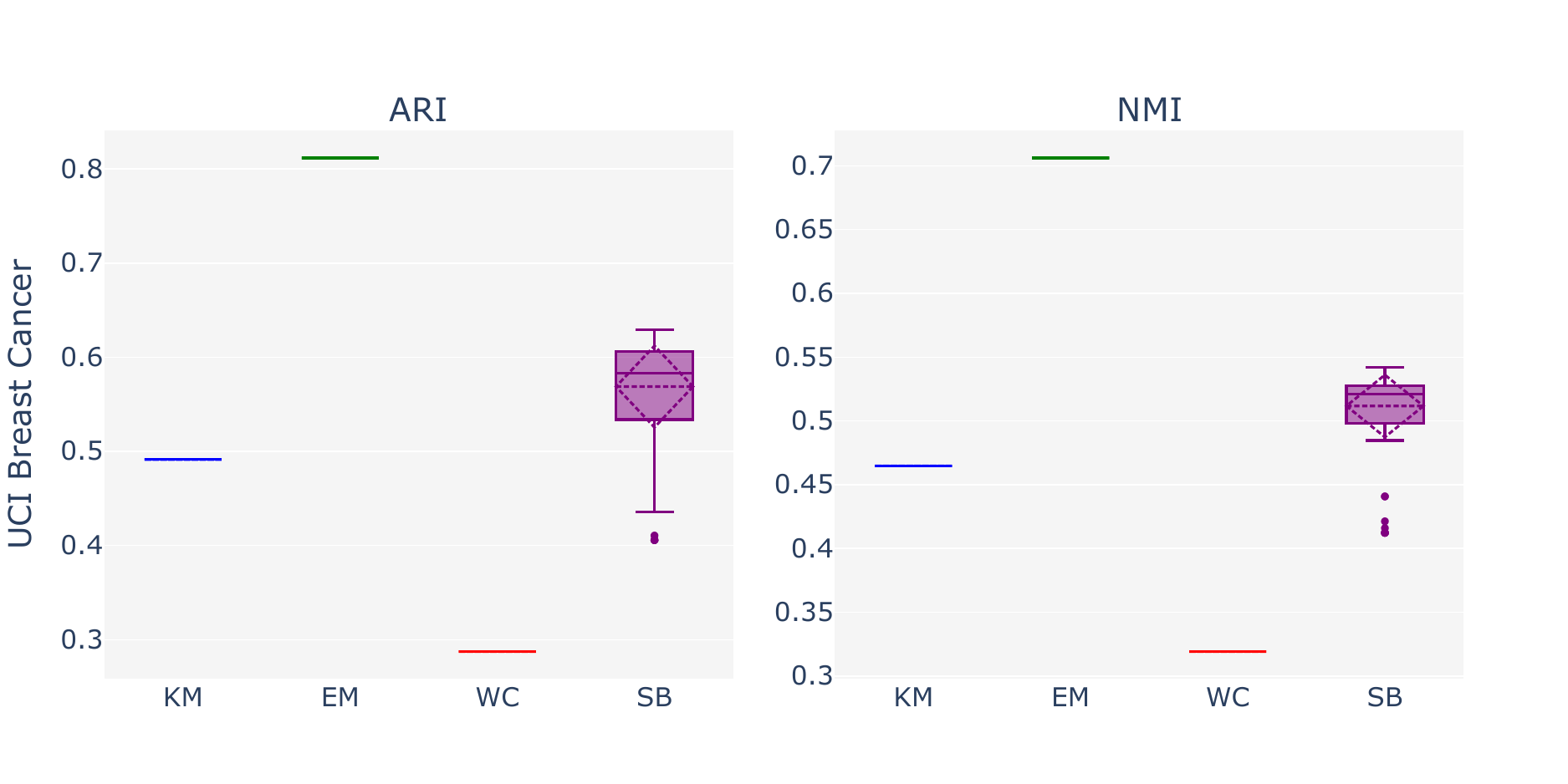}

}

\caption{\label{fig-cancer-scores}ARI and NMI scores of k-means++
(blue), EM (green), Ward Clustering (red), and Spectral Bridges (purple)
on the UCI Breast Cancer dataset.}

\end{figure}

The Spectral Bridges algorithm was also put to the test against the same
competitors using scikit-learn's UCI Breast Cancer data. Once again,
this new method performs well although the advantage is not as obvious
in this case (see Figure~\ref{fig-cancer-scores}). However, in none of
our tests has it ranked worse than k-means++. The results are displayed
as a boxplot generated from 200 iterations of each algorithm using a
different seed, in order to better grasp the variability lying in the
seed dependent nature of the k-means++, Expectation Maximization and
Spectral Bridges algorithms.

\begin{figure}

\begin{minipage}{0.25\linewidth}

\includegraphics{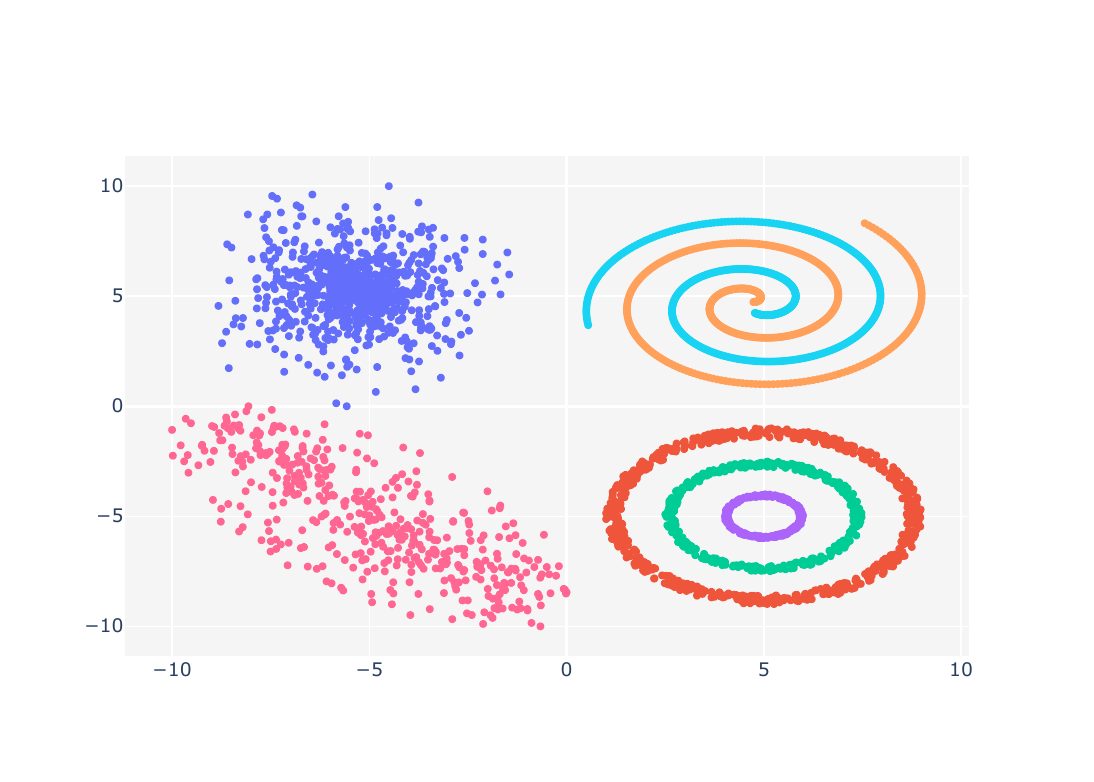}

\subcaption{\label{}Impossible}
\end{minipage}%
\begin{minipage}{0.25\linewidth}

\includegraphics{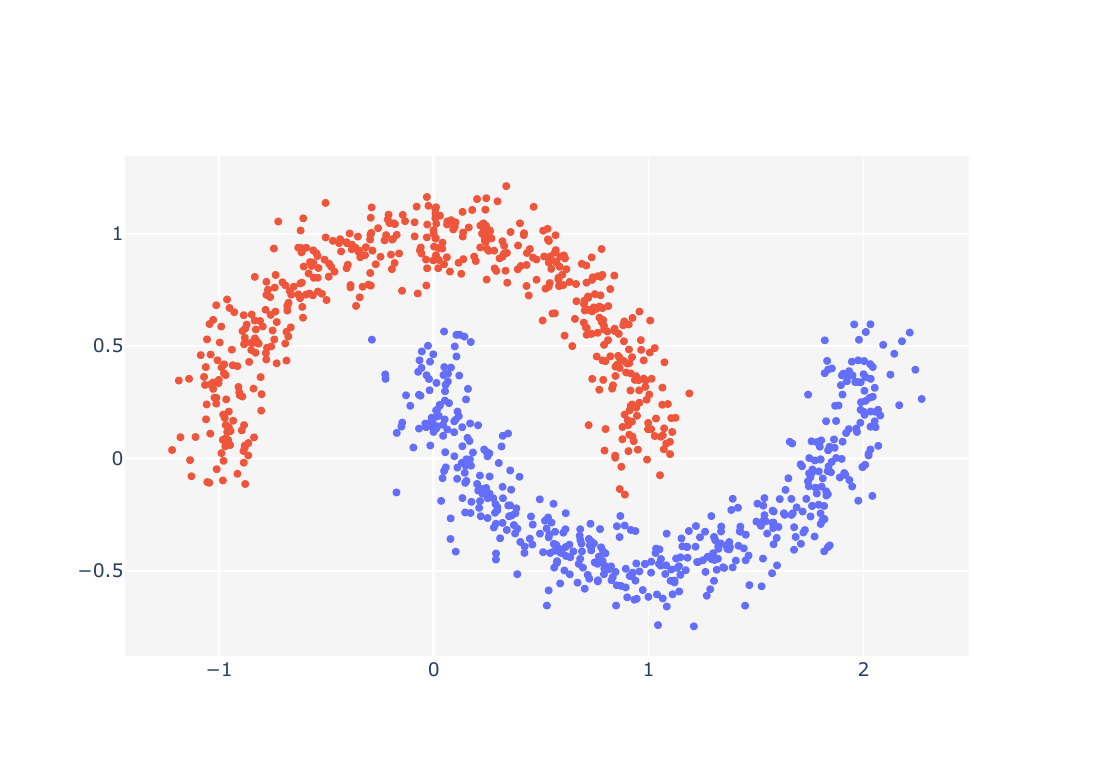}

\subcaption{\label{}Moons}
\end{minipage}%
\begin{minipage}{0.25\linewidth}

\includegraphics{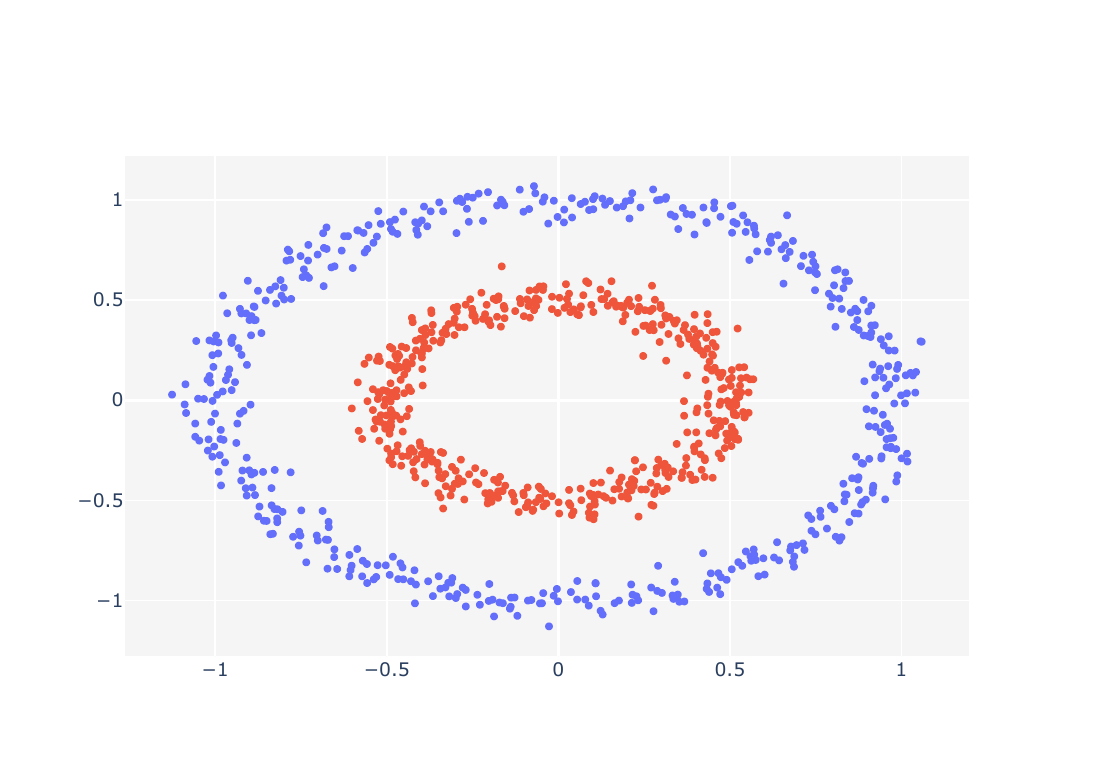}

\subcaption{\label{}Circles}
\end{minipage}%
\begin{minipage}{0.25\linewidth}

\includegraphics{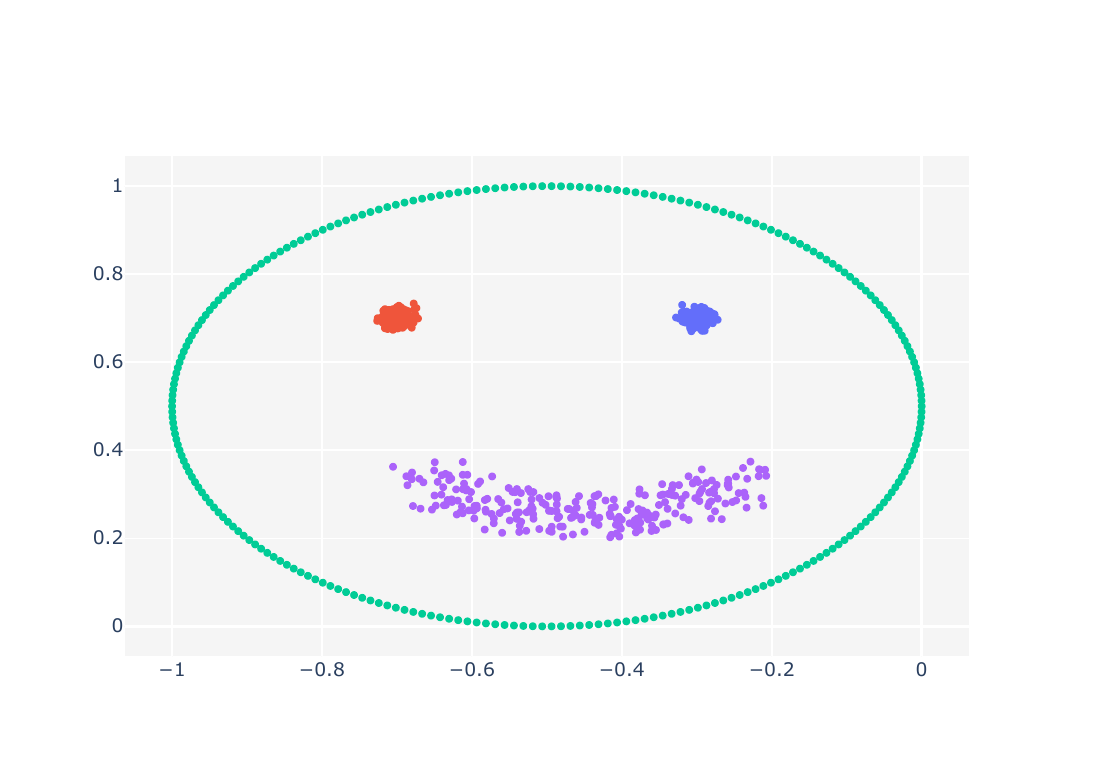}

\subcaption{\label{}Smile}
\end{minipage}%

\caption{\label{fig-toy-datasets}Four toy datasets.}

\end{figure}

Since the Spectral Bridges algorithm is expected to excel at discerning
complex and intricate cluster structures, an array of four toy datasets
was collected, as illustrated in Figure~\ref{fig-toy-datasets}.

Multiple algorithms, including the proposed one, were benchmarked in the
exact same manner as for the UCI Breast Cancer data. The results show
that the proposed method outperforms all tested algorithms (DBSCAN,
k-means++, Expectation Maximization, and Ward Clustering) while
requiring few hyperparameters. As previously discussed, DBSCAN's
parameters were optimized using the ground truth labels to represent a
best-case scenario; however, in practical applications, suboptimal
performance is more likely. Despite this optimization, the
Spectral-Bridge algorithm still demonstrates superior ability to capture
and represent the underlying cluster structures.

\begin{figure}

\centering{

\includegraphics{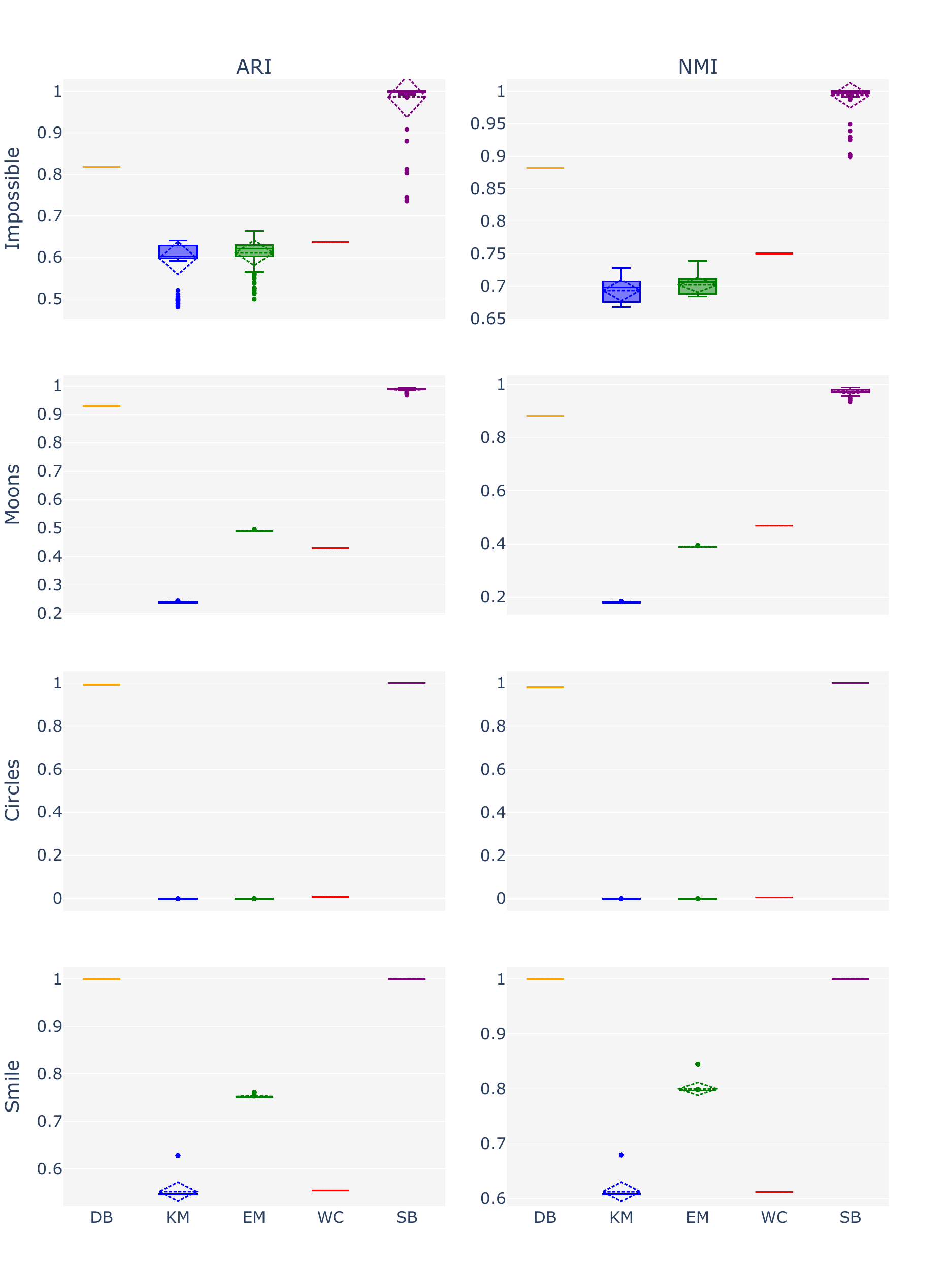}

}

\caption{\label{fig-synthetic-scores}ARI and NMI scores of Spectral
Bridges and competitors on standard synthetic toy datasets.}

\end{figure}%

\subsection{Noise robustness}\label{noise-robustness}

\begin{figure}

\begin{minipage}{0.33\linewidth}

\includegraphics{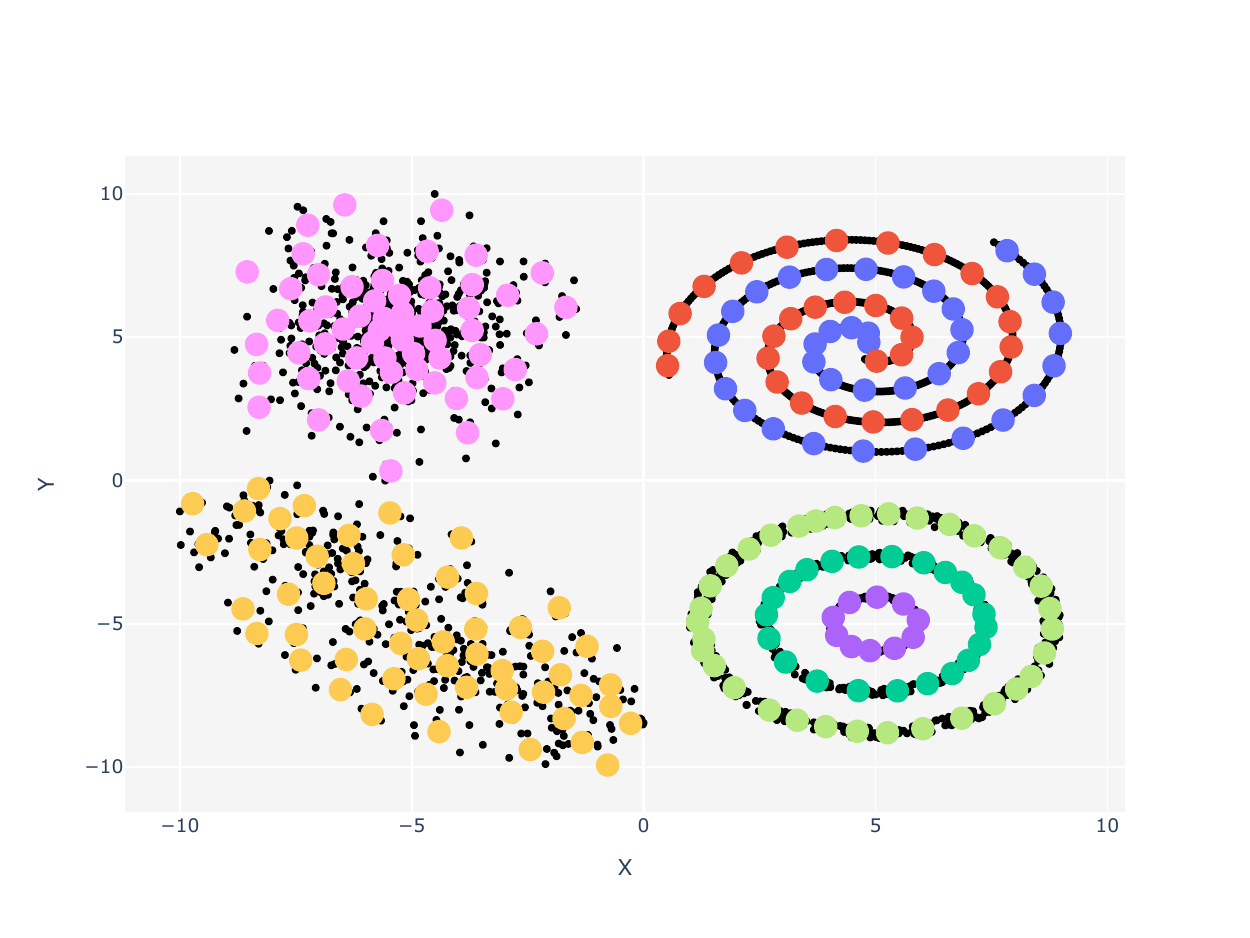}

\subcaption{\label{}Clean}
\end{minipage}%
\begin{minipage}{0.33\linewidth}

\includegraphics{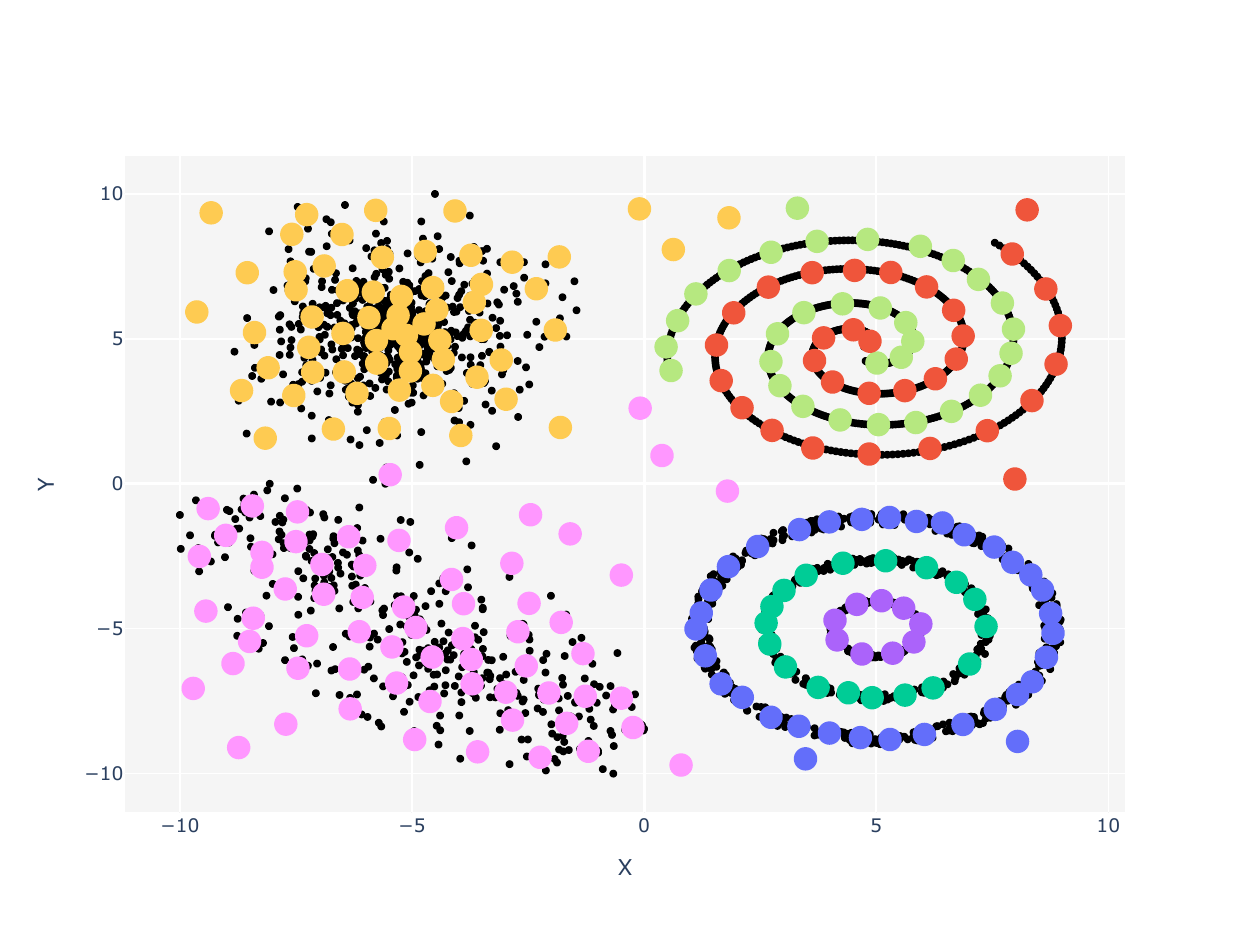}

\subcaption{\label{}Uniform noise}
\end{minipage}%
\begin{minipage}{0.33\linewidth}

\includegraphics{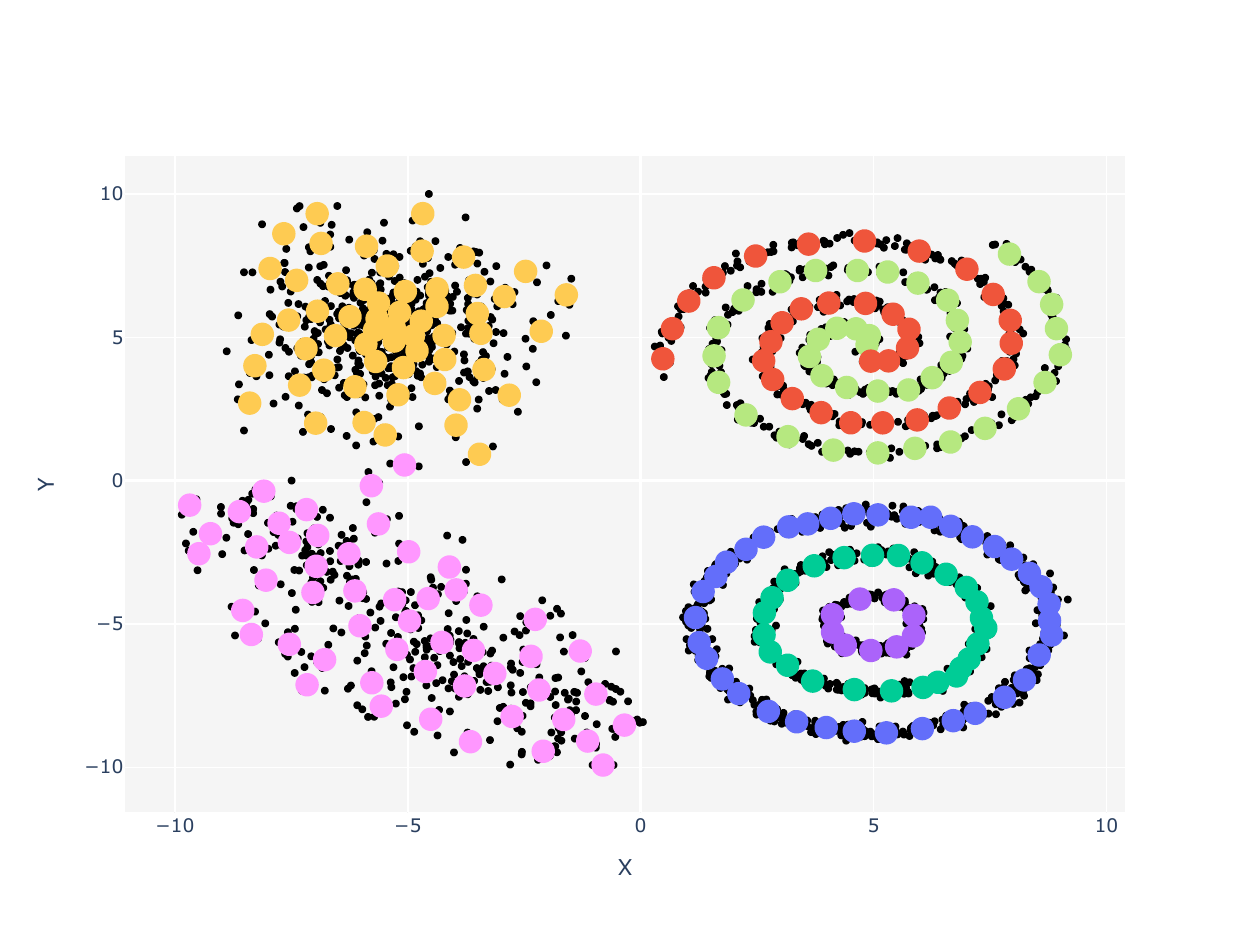}

\subcaption{\label{}Gaussian noise}
\end{minipage}%

\caption{\label{fig-noise-robustness}Three representations of the
algorithm's predicted cluster centers are displayed as colored dots,
with each point of the Impossible dataset shown as a small black dot. In
the left graph, the dataset is unmodified. In the center graph, 250
uniformly distributed samples were added. In the right graph, Gaussian
noise perturbations with \(\sigma = 0.1\) were applied.}

\end{figure}

To evaluate the noise robustness of the algorithm, two experimental
setups were devised: one involved introducing Gaussian-distributed
perturbations to the data, and the other involved concatenating
uniformly distributed points within a predefined rectangular region
(determined by the span of the dataset) to the existing dataset. As
illustrated in Figure~\ref{fig-noise-robustness}, the tests demonstrate
that in both scenarios, the algorithm exhibits a high degree of
insensitivity to noise.

\subsection{Hyperparameter values effect on
accuracy}\label{hyperparameter-values-effect-on-accuracy}

\begin{figure}

\centering{

\includegraphics{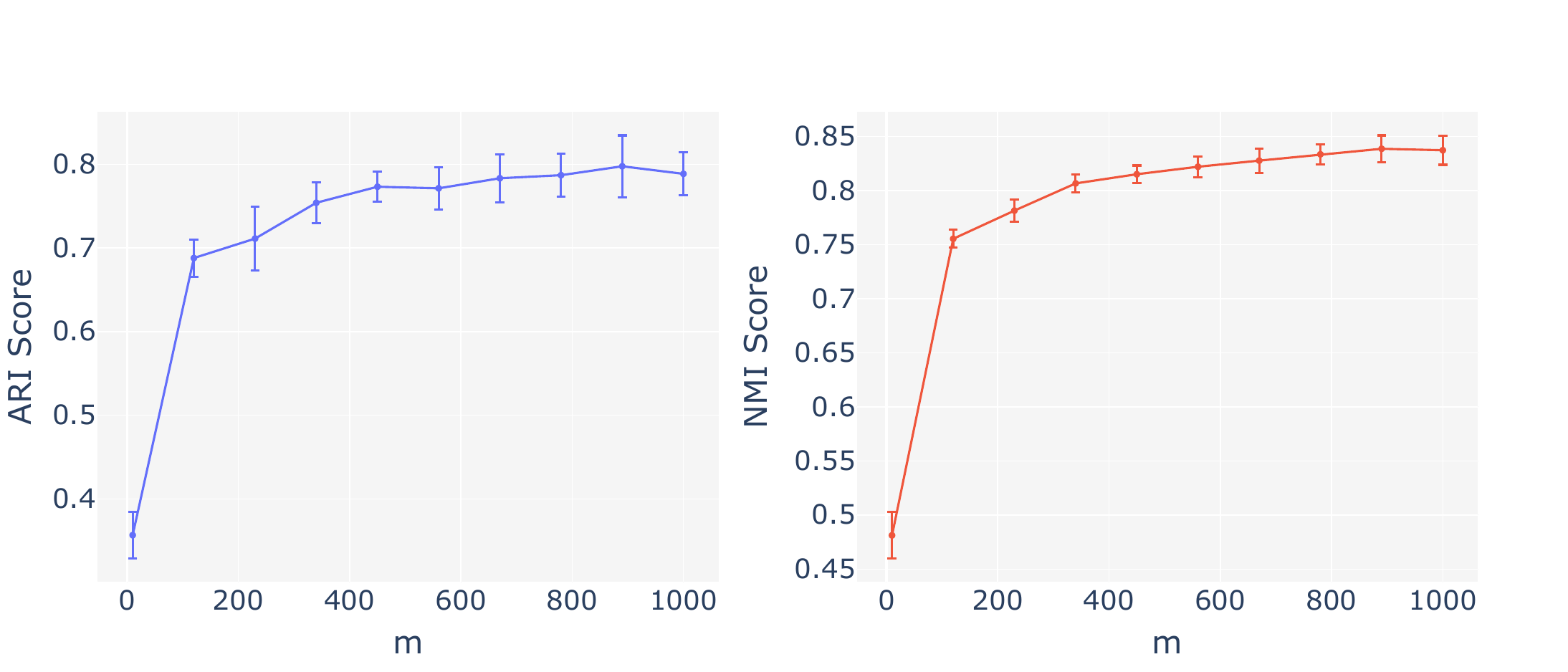}

}

\caption{\label{fig-m-vs-score}ARI and NMI scores of Spectral Bridges
with varying values of \(m\).}

\end{figure}

To better understand and measure the significance of choosing the right
values for the hyperparameters of the proposed algorithm, that it to say
the number of Voronoï regions \(m\), Spectral Bridges was run on the PCA
\(h = 32\) embedded MNIST dataset with varying values of
\(m \in \{10, 120, 230, 340, 450, 560, 670, 780, 890, 1000 \}\). The
case \(m = 10\) is equivalent to the k-means++ algorithm. ARI and NMI
scores are recorded over 20 consecutive iterations and subsequently
plotted. As shown by Figure~\ref{fig-m-vs-score}, the accuracy seems to
be consistently increasing with values of \(m\), although the largest
observed gap occurs between values of \(m = 10\) and \(m = 120\),
indicating a tremendous improvement over the classical k-means++
framework even for empirically suboptimal hyperparameter values.

\section{Conclusive remarks}\label{conclusive-remarks}

Spectral Bridges is an original clustering algorithm which presents a
novel approach by integrating the strengths of traditional k-means and
spectral clustering frameworks. This algorithm utilizes a simple
affinity measure for spectral clustering, which is derived from the
minimal margin between pairs of Voronoï regions.

The algorithm demonstrates scalability, handling large datasets
efficiently through a balanced computational complexity between the
k-means clustering and eigen-decomposition steps. As a non-parametric
method, Spectral Bridges does not rely on strong assumptions about data
distribution, enhancing its versatility across various data types. It
performs exceptionally well with both synthetic and real-world data and
consistently outperforms conventional clustering algorithms such as
k-means, DBSCAN, and mixture models.

The design of Spectral Bridges ensures robustness to noise, a
significant advantage in real-world applications. Additionally, the
algorithm requires minimal hyperparameters, primarily the number of
Voronoï regions, making it straightforward to tune and deploy.

Furthermore, Spectral Bridges can be kernelized, allowing it to handle
data in similarity space directly, which enhances its flexibility and
applicability. Overall, Spectral Bridges is a powerful, robust, and
scalable clustering algorithm that offers significant improvements over
traditional methods, making it an excellent tool for advanced clustering
tasks across numerous domains.

\section{Appendix}\label{appendix}

\subsection{Derivation of the bridge affinity}\label{gain}

We denote a bridge as a segment connecting two centroids
\(\boldsymbol \mu_k\) and \(\boldsymbol \mu_l\). The inertia of a bridge
between \(\mathcal{V}_k\) and \(\mathcal{V}_l\) is defined as \[
B_{kl} = \sum_{\boldsymbol x_i\in \mathcal{V}_k \cup \mathcal{V}_l} \|\boldsymbol x_i - \boldsymbol p_{kl}(\boldsymbol x_i)\|^2,
\] where \[
\boldsymbol p_{kl}(\boldsymbol x_i) = \boldsymbol \mu_{k} + t_i(\boldsymbol \mu_{l} - \boldsymbol \mu_{k}),
\] with \[
t_i  = \min\left(1, \max\left(0, \frac{\langle \boldsymbol x_i - \boldsymbol \mu_k | \boldsymbol \mu_l - \boldsymbol \mu_k\rangle}{\|  \boldsymbol \mu_l - \boldsymbol \mu_k \|^2}\right)\right). 
\]

\(B_{kl}\), the bridge inertia between centroids \(k\) and \(l\), can be
expressed as the sum of three terms, which represents the projection
onto each centroïds and onto the segment:

\[
\begin{aligned}
B_{kl} &=& \sum_{i \mid t_i=0} \|\boldsymbol x_i - \boldsymbol \mu_k\|^2  + \sum_{i \mid t_i=1} \|\boldsymbol x_i - \boldsymbol \mu_l\|^2 + \sum_{i \mid t_i\in ]0,1[} \|\boldsymbol x_i - \boldsymbol p_{kl}(\boldsymbol x_i)\|^2.
\end{aligned}
\]

The last term may be decomposed in two parts corresponding to the points
of the two Voronoï regions which are projected on the segment:

\[
\begin{aligned}
\sum_{i \mid t_i\in ]0,1[} \|\boldsymbol x_i - \boldsymbol p_{kl}(\boldsymbol x_i)\|^2 &= &\sum_{i \mid t_i\in ]0,\frac{1}{2}[} \|\boldsymbol x_i - \boldsymbol p_{kl}(\boldsymbol x_i)\|^2 + \sum_{i \mid t_i\in [\frac{1}{2},1[} \|\boldsymbol x_i - \boldsymbol p_{kl}(\boldsymbol x_i)\|^2\\
\end{aligned}
\] and each part further decomposed using Pythagore \[
\begin{aligned}
\sum_{i \mid t_i\in ]0,\frac{1}{2}[} \|\boldsymbol x_i - \boldsymbol p_{kl}(\boldsymbol x_i)\|^2 &=& \sum_{i \mid t_i\in ]0,\frac{1}{2}[} \|\boldsymbol x_i - \boldsymbol \mu_k\|^2 - \sum_{i \mid t_i\in ]0,\frac{1}{2}[} \|\boldsymbol \mu_k - \boldsymbol p_{kl}(\boldsymbol x_i)\|^2\\
&=& \sum_{i \mid t_i\in ]0,\frac{1}{2}[} \|\boldsymbol x_i - \boldsymbol \mu_k\|^2 - \sum_{i \mid t_i\in ]0,\frac{1}{2}[} \|t_i (\boldsymbol \mu_k - \boldsymbol \mu_{l})\|^2,
\end{aligned}
\]

\[
\begin{aligned}
\sum_{i \mid t_i\in ]\frac{1}{2},1[} \|\boldsymbol x_i - \boldsymbol p_{kl}(\boldsymbol x_i)\|^2 &=& \sum_{i \mid t_i\in ]0,\frac{1}{2}[} \|\boldsymbol x_i - \boldsymbol \mu_l\|^2 - \sum_{i \mid t_i\in ]0,\frac{1}{2}[} \|\boldsymbol \mu_l - \boldsymbol p_{kl}(\boldsymbol x_i)\|^2\\
&=& \sum_{i \mid t_i\in ]\frac{1}{2},1[} \|\boldsymbol x_i - \boldsymbol \mu_k\|^2 - \sum_{i \mid t_i\in ]0,\frac{1}{2}[} \|(1-t_i) (\boldsymbol \mu_k - \boldsymbol \mu_{l})\|^2
\end{aligned}
\]

Thus \[
\begin{aligned}
B_{kl}- I_{kl} &=&  \sum_{i \mid t_i\in ]0,\frac{1}{2}[} t_i^2 \|\boldsymbol \mu_k - \boldsymbol \mu_l\|^2 + \sum_{i \mid t_i\in ]\frac{1}{2},1[} (1-t_i)^2 \|\boldsymbol \mu_k - \boldsymbol \mu_l\|^2,\\
\frac{B_{kl}- I_{kl}}{\|\boldsymbol \mu_k - \boldsymbol \mu_l\|^2} &=& \sum_{i \mid t_i\in ]0,\frac{1}{2}[} t_i^2  + \sum_{i \mid t_i\in ]\frac{1}{2},1[} (1-t_i)^2, \\
\frac{B_{kl}- I_{kl}}{(n_k+n_l)\|\boldsymbol \mu_k - \boldsymbol \mu_l\|^2} &=& \frac{\sum_{\boldsymbol{x_i} \in \mathcal V_k} \langle \boldsymbol{x_i} - \boldsymbol{\mu}_k \vert \boldsymbol{\mu}_l - \boldsymbol{\mu}_k \rangle_+^2  \sum_{\boldsymbol{x_i} \in \mathcal V_l} \langle \boldsymbol{x_i} - \boldsymbol{\mu}_l \vert \boldsymbol{\mu}_k - \boldsymbol{\mu}_l\rangle_+^2}{(n_k+n_l)\|\boldsymbol \mu_k - \boldsymbol \mu_l\|^4}.
\end{aligned}
\]

\subsection{Code}\label{code}

\subsubsection{Implementation}\label{implementation}

Numerical experiments have been conducted in Python. The python scripts
to reproduce the simulations and figures are available at
\url{https://github.com/flheight/Spectral-Bridges}. The Spectral Bridge
algorithm is implemented both in

\begin{itemize}
\tightlist
\item
  Python: \url{https://pypi.org/project/spectral-bridges}, and
\item
  R: \url{https://github.com/cambroise/spectral-bridges-Rpackage}.
\end{itemize}

\subsubsection{Affinity matrix
computation}\label{affinity-matrix-computation}

Taking a closer look at the second step of
 Algorithm~\ref{alg-spectral-bridges} , that is the affinity matrix
calculation with a \(O(n \times m \times d)\) time complexity, most
operations can be parallelized leaving a single loop, bundling together
\(m^2\) dot products into only \(m\) matrix multiplications, thus
allowing for an efficient construction in both high and low level
programming languages. Though the complexity of the algorithm remains
unchanged, libraries such as Basic Linear Algebra Subprograms can render
the calculations orders of magnitude faster. Moreover, the symmetrical
nature of the bridge affinity can be used to effectively halve the
computation time.

The calculation of the affinity matrix is highlighted by the Python code
Listing~\ref{lst-code-affinity}. Though it could be even more optimized,
the following code snippet is approximately 200 times faster than a
naive implementation on a small dataset comprised of \(n = 3594\),
\(d = 2\) points, and a value of \(m = 250\).

Notice that the Python code is significantly faster than the R code.

\begin{codelisting}

\caption{\label{lst-code-affinity}Python code for affinity matrix
computation}

\centering{

\begin{Shaded}
\begin{Highlighting}[]
\CommentTok{\# Initialize the affinity matrix}
\NormalTok{affinity }\OperatorTok{=}\NormalTok{ np.empty((}\VariableTok{self}\NormalTok{.n\_nodes, }\VariableTok{self}\NormalTok{.n\_nodes))}

\CommentTok{\# Center each Voronoi region around its centroid}
\NormalTok{X\_centered }\OperatorTok{=}\NormalTok{ [}
\NormalTok{    X[kmeans.labels\_ }\OperatorTok{==}\NormalTok{ i] }\OperatorTok{{-}}\NormalTok{ kmeans.cluster\_centers\_[i] }\ControlFlowTok{for}\NormalTok{ i }\KeywordTok{in} \BuiltInTok{range}\NormalTok{(}\VariableTok{self}\NormalTok{.n\_nodes)}
\NormalTok{]}

\CommentTok{\# Count the total number of points in each pair of regions}
\NormalTok{counts }\OperatorTok{=}\NormalTok{ np.array([X\_centered[i].shape[}\DecValTok{0}\NormalTok{] }\ControlFlowTok{for}\NormalTok{ i }\KeywordTok{in} \BuiltInTok{range}\NormalTok{(}\VariableTok{self}\NormalTok{.n\_nodes)])}
\NormalTok{counts }\OperatorTok{=}\NormalTok{ counts[np.newaxis, :] }\OperatorTok{+}\NormalTok{ counts[:, np.newaxis]}

\CommentTok{\# Compute the segments between each pair of centroids and their squared Euclidean norm}
\NormalTok{segments }\OperatorTok{=}\NormalTok{ (}
\NormalTok{    kmeans.cluster\_centers\_[np.newaxis, :] }\OperatorTok{{-}}\NormalTok{ kmeans.cluster\_centers\_[:, np.newaxis]}
\NormalTok{)}
\NormalTok{dists }\OperatorTok{=}\NormalTok{ np.einsum(}\StringTok{"ijk,ijk{-}\textgreater{}ij"}\NormalTok{, segments, segments)}
\NormalTok{np.fill\_diagonal(dists, }\DecValTok{1}\NormalTok{)  }\CommentTok{\# Avoid dividing by zero}

\CommentTok{\# Assign each row of the affinity matrix}
\ControlFlowTok{for}\NormalTok{ i }\KeywordTok{in} \BuiltInTok{range}\NormalTok{(}\VariableTok{self}\NormalTok{.n\_nodes):}
\NormalTok{    projs }\OperatorTok{=}\NormalTok{ np.maximum(np.dot(X\_centered[i], segments[i].T), }\DecValTok{0}\NormalTok{)}
\NormalTok{    affinity[i] }\OperatorTok{=}\NormalTok{ np.einsum(}\StringTok{"ij,ij{-}\textgreater{}j"}\NormalTok{, projs, projs)}

\CommentTok{\# Symmetrize the matrix and normalize, as well as taking the element{-}wise square root}
\NormalTok{affinity }\OperatorTok{=}\NormalTok{ np.sqrt(affinity }\OperatorTok{+}\NormalTok{ affinity.T) }\OperatorTok{/}\NormalTok{ (np.sqrt(counts) }\OperatorTok{*}\NormalTok{ dists)}
\NormalTok{affinity }\OperatorTok{{-}=} \FloatTok{0.5} \OperatorTok{*}\NormalTok{ affinity.}\BuiltInTok{max}\NormalTok{()  }\CommentTok{\# For numerical stability}

\CommentTok{\# Apply the exponential transformation}
\NormalTok{q10, q90 }\OperatorTok{=}\NormalTok{ np.quantile(affinity, [}\FloatTok{0.1}\NormalTok{, }\FloatTok{0.9}\NormalTok{])}

\NormalTok{gamma }\OperatorTok{=}\NormalTok{ np.log(}\VariableTok{self}\NormalTok{.M) }\OperatorTok{/}\NormalTok{ (q90 }\OperatorTok{{-}}\NormalTok{ q10)}
\NormalTok{affinity }\OperatorTok{=}\NormalTok{ np.exp(gamma }\OperatorTok{*}\NormalTok{ affinity)}
\end{Highlighting}
\end{Shaded}

}

\end{codelisting}%

\section*{References}\label{references}
\addcontentsline{toc}{section}{References}

\phantomsection\label{refs}


%
%
%
%
%
%

\end{document}